\documentclass[reprint,aps,pra]{revtex4-2}
\usepackage{dcolumn}
\usepackage[colorlinks=true, pdfstartview=FitV, linkcolor=red, citecolor=blue, urlcolor=blue]{hyperref}
\usepackage{graphicx,color}
\usepackage{bm}
\usepackage{amsmath}
\usepackage{diagbox} 
\usepackage{hhline}

\newcommand\nn{\nonumber \\}

\newcommand\bk{{\bm k}}

\newcommand\bd{{\bm d}}
\newcommand\bp{{\bm p}}
\newcommand\bx{{\bm x}}

\newcommand\bq{{\bm q}}
\newcommand\br{{\bm r}}
\newcommand\rd{{\rm d}}

\newcommand{\lk}{\left(}
\newcommand{\rk}{\right)}
\newcommand{\ltk}{\left\{}
\newcommand{\rtk}{\right\}}
\newcommand{\ldk}{\left[}

\newcommand{\rdk}{\right]}
\newcommand\beq{ \begin{eqnarray} }
\newcommand\eeq{ \end{eqnarray} }

\begin{document}
\title{Two-body problem of impurity atoms in dipolar Fermi gas}
\author{E. Nakano}
\affiliation{Department of Mathematics and Physics, Kochi University, Kochi 780-8520, Japan}
\author{T. Miyakawa}
\affiliation{Faculty of Education, Aichi University of Education, Japan}
\author{H. Yabu}
\affiliation{Department of Physics, Ritsumeikan University, Siga, Japan}
\date{\today}
\begin{abstract} 
The polarized dipolar Fermi gas shows exotic properties at low temperatures, 
characterized by an axially-deformed Fermi surface and anisotropic single-particle energy, 
due to the long-range and anisotropic nature of dipole-dipole interaction. 
In cold-atom experiments such a system has been realized, e.g., in degenerate gas of Er and Dy atoms. 
In the case that non-dipolar impurity atoms are introduced in such system, 
they undergoes an induced interaction mediated by the density fluctuations of the background dipolar Fermi gas. 
We derive the induced interaction potential to the single-loop order of fluctuations 
and show that it becomes indeed an anisotropic Ruderman-Kittel-Kasuya-Yosida-type potential 
which preserves the axial symmetry around the polarization axis. 
We then solve the two-body problem of impurity atoms interacting 
via the anisotropic potential and figure out the dependence of bound state and scattering properties  
on the parameters of dipolar Fermi gas. 
\end{abstract}

\maketitle

\section{Introduction} 
Quantum many-body system with dipole-dipole interaction (DDI), 
due to its long range and anisotropic nature, 
is expected to show a variety of interesting phases 
which include 
supersolid states \cite{Kadau_2016,Tanzi_2019,Hertkorn_2021}, 
quantum droplets \cite{Ferrier_Barbut_2016}, and 
unconventional magnetism \cite{Fregoso_2009b,Fregoso_2010,Sogo_2012,Bhongale_2013} 
\& superfluids \cite{Baranov_2002,Baranov_2004,Bruun_2008,Cooper_2009,Levinsen_2011}. 
In cold-atom experiments, 
the quantum many-body system of dipolar atoms has been recently realized 
using the atoms of Er and Dy isotopes with large magnetic dipole moments, 
which are trapped and cooled down to be
Bose-Einstein condensates (BEC) \cite{Griesmaier_2005,Lu_2011,Aikawa_2012} or 
Fermi-degenerate states \cite{Lu_2012,Aikawa_2014a,Naylor_2015}.  
Such systems provide the platform to investigate the above-mentioned exotic phases caused by DDI experimentally 
in precise and well controlled ways \cite{Baier_2018,Durastante_2020,Chomaz_2022}. 

In dipolar Bose-atom systems, 
the attractive part of DDI can induce the collapse of BEC 
or the instability toward the formation of a finite momentum condensate, 
which is triggered by a softening of roton excitation 
as a precursor of supersolidity \cite{Kadau_2016,Hertkorn_2021,Klaus_2022,Scheiermann_2023}. 
Such systems of dipolar Bose gas with the long-range attraction 
can be sustained by the quantum corrections involving a short-range repulsion, 
which appears as a higher-order density term in the energy functional \cite{Petrov_2015}. 
This sort of mechanism works also on the stabilization of quantum droplets \cite{Ferrier_Barbut_2016,Kumar_2017}. 

In fermionic systems, on the other hand, 
focusing on the degenerate dipolar Fermi gas 
in which dipoles are polarized along a specific direction by an external field 
at low temperatures, 
theoretical studies so far 
have predicted that the Fermi surface undergoes ellipsoidal deformations at weak DDIs 
\cite{Miyakawa_2008,Sogo_2009,Tohyama_2009,Zhang_2010,Parish_2012,Zhou_2021}. 
As a consequence of the phase-space deformation, 
collective excitations, e.g., zero sound, propagate anisotropically, 
and exhibit the angular dependence with respect to the direction of dipole polarization  
\cite{Chan_2010,Ronen_2010,Sieberer_2011,MNY_2023}. 
In experiments, the Fermi-surface deformation has actually been observed 
using a trapped erbium gas ($^{167}$Er) \cite{Aikawa_2014b}, 
and such a polarized dipolar Fermi gas with a uniaxial symmetry provides 
a quantum analog of nematic liquid crystals \cite{Fregoso_2009}, 
in which transport or thermalization properties 
exhibit similar anisotropy with classical liquid crystals \cite{Aikawa_2014c,Wang_2021}. 

Another interesting many-body system with DDI is 
gaseous mixtures of dipolar Fermi atoms and non-dipolar atoms. 
This kind of systems is feasible in experimental setups, 
for instance, using gaseous mixtures of Dy \& K 
\cite{Ravensbergen_2018,Ravensbergen_2020,Ye_2022}, 
Cr \& Li  \cite{Ciamei_2022,Ciamei_2022_2}, 
and Er \& Li atoms 
\cite{Sch_fer_2022,Sch_fer_2023,Sch_fer_2023_2}. 
So far, theoretical studies of this kind of mixture have been done about  
the superfluidity with dipolar and non-dipolar atom pairing \cite{baarsma2017fermi} 
and the density-density correlation functions for zero sounds \cite{MNY_2023}.  

In the present paper, we study the mixture at zero temperature 
in which the non-dipolar atomic gas is so dilute as to be impurities 
immersed in the degenerate dipolar Fermi gas. 
Quasiparticle property of such an impurity atom has been investigated previously 
to find that its dispersion relation shows anisotropy attributed to the deformation of Fermi surface \cite{Nishimura_2021}. 
The quasiparticle is a variant of so-called Fermi polarons, 
which have been actively studied in a recent decade 
in cold-atom systems  \cite{Chevy_2010,Massignan_2014,Schmidt_2018,Tajima_2021,Scazza_2022}. 
In Fermi-polaron problems, while one usually looks into single-particle properties, 
in this work we focus on the two-body effective interaction 
between non-dipolar impurity atoms induced by density fluctuations of the medium of dipolar Fermi gas, 
and investigate the two-body problem of the impurity atoms.
It will provide the basis for further studies in understanding of few-body correlations of the minority particles 
in the medium of majority fermions \cite{Camacho_Guardian_2018,Moriya_2021,Tajima_2022,Muir_2022}. 

In section II in the paper, we present the effective Hamiltonian 
for the mixture of dipolar Fermi atoms and non-dipolar impurity atoms, 
in which the medium dipolar Fermi gas with deformed Fermi surface 
is described in Hartree-Fock approximation 
and a contact interaction between medium and impurity atoms is given 
in terms of s-wave scattering length. 
In section III, we derive the induced interaction potential between two impurity atoms 
using the density-density correlation function of medium dipolar fermions. 
The potential is found to be that of anisotropic Ruderman-Kittel-Kasuya-Yosida (RKKY) type 
\cite{Ruderman_1954,Kasuya_1956,Yoshida_1957}.  
In section IV, we solve Schr\"{o}dinger equation with the RKKY potential 
to figure out the conditions for two-body bound states to emerge. 
In section V, we evaluate the two-impurity atom scattering amplitude in Born approximation, 
and observe how its magnitude depends on 
the directions of the initial and final momenta and also the dipole polarization. 
We also make the partial-wave analysis 
to find the angular momentum mixing and the dominant channels 
in low energy scattering processes. 
The last section VI is devoted to summary and outlooks. 
Throughout the paper we use the natural units where $\hbar=1$, 
$\mu_0=1$ for the vacuum magnetic permeability, 
and the volume of the system is fixed to be unity. 
We also use the abbreviated notations for integrals in real and momentum spaces as  
$\sum_\bx \equiv \int {\rd}^3\bx$ and $\sum_\bq \equiv \int {\rd}^3\bq/(2\pi)^3$. 
\section{Effective Hamiltonian}
We consider the spatially-homogeneous mixture of a atomic Fermi gas 
with a large dipole moment and non-dipolar impurity atoms at zero temperature, 
in which all dipole moments are polarized in a specific direction. 
For a while we do not specify the particle statistics of impurity atoms. 
The model Hamiltonian of the system is given by 
\beq
H&=&
\sum_\bk \epsilon_{1\bk} c_{1\bk}^\dagger c_{1\bk}
+
\sum_\bk \epsilon_{2\bk} c_{2\bk}^\dagger c_{2\bk}
\nn
&&+
\frac{1}{2}
\sum_{\bk,\bk',\bq} V_{dd}(\bq) c_{1\bk}^\dagger c_{1\bk'+\bq}^\dagger c_{1\bk'} c_{1\bk+\bq}
\nn
&&+
g \sum_{\bk,\bk',\bq}  c_{1\bk}^\dagger c_{2\bk'+\bq}^\dagger c_{2\bk'} c_{1\bk+\bq}
\label{Hamil1}
\eeq
where $c_{1\bk}$ and $c_{1\bk}^\dagger$ are 
canonical annihilation and creation operators of 
dipolar fermions of the mass $m_1$ and the spatial momentum $\bk$, 
and the single-particle energy $\epsilon_{1\bk}=\bk^2/2m_1$. 
Similarly,  $c_{2\bk}$ and $c_{2\bk}^\dagger$ are annihilation and creation operators of 
non-dipolar impurity atoms of the momentum $\bk$ and the mass $m_2$, 
and the single-particle energy $\epsilon_{2\bk}=\bk^2/2m_2$. 

The third term in (\ref{Hamil1}) is
the dipole-dipole interaction, 
the potential of which is
\beq
V_{dd}(\bq) &=& \frac{4\pi}{3} \bd^2\lk3\cos^2\theta_\bq-1\rk
\eeq
where $\bd$ is the (magnetic) dipole moment, 
and $\theta_\bq$ is the angle between $\bq$ and $\bd$. 
The last term is the contact interaction between dipolar and non-dipolar atoms; 
the interaction strength is given by 
$g=2\pi a_{12}/m_{12}$ 
with $s$-wave scattering length $a_{12}$ and 
the reduced mass $m_{12}=m_1m_2/(m_1+m_2)$. 

Now we construct an effective Hamiltonian based on 
the self-consistent Hartree-Fock (HF) approximation for degenerate dipolar fermions,  
as in the previous works \cite{Miyakawa_2008,Sogo_2009,Ronen_2010}. 
Introducing  
the annihilation and creation operators, 
$a_{\bk}$ and $a_{\bk}^\dagger$ for the particle mode and 
$b_{\bk}$ and $b_{\bk}^\dagger$ for the hole mode, 
for the dipolar fermions
with the HF single-particle energy $\epsilon_{\bk}$, 
the operators $c_{1\bk}$ and $c_{1\bk}^\dagger$ can be rewritten as
\beq
c_{1\bk} &=& \theta\lk \epsilon_{\bk} -\epsilon_F\rk a_{\bk} 
+ \theta\lk \epsilon_F-\epsilon_{\bk}\rk b^\dagger_{-\bk}, 
\nn
c_{1\bk}^\dagger &=& \theta\lk \epsilon_{\bk} -\epsilon_F\rk a_{\bk}^\dagger 
+ \theta\lk \epsilon_F-\epsilon_{\bk}\rk b_{-\bk}, 
\label{newbase1}
\eeq
where $\epsilon_{F}$ is the Fermi energy. 
Consequently, 
the effective Hamiltonian in the HF approximation becomes
\beq
H_{\rm HF}&=& E_0 +
\sum_\bk \epsilon_{2\bk} c_{2\bk}^\dagger c_{2\bk}
\nn
&&+
\sum_\bk  \theta\lk \epsilon_{\bk} -\epsilon_F\rk \epsilon_{\bk} a_{\bk}^\dagger a_{\bk}
-
\sum_\bk \theta\lk \epsilon_F-\epsilon_{\bk} \rk \epsilon_{\bk} b_{\bk}^\dagger b_{\bk}
\nn
&&+
\frac{1}{2}\sum_{\bk,\bk',\bq} 
V_{dd}(\bq) 
{\mathcal N}\ldk c_{1\bk}^\dagger c_{1\bk'+\bq}^\dagger c_{1\bk'} c_{1\bk+\bq}\rdk 
\nn
&&+
g \sum_{\bk,\bk',\bq}  
{\mathcal N}\ldk c_{1\bk}^\dagger c_{2\bk'+\bq}^\dagger c_{2\bk'} c_{1\bk+\bq}\rdk 
\eeq
where $E_0$ is the Hartree rest energy, and 
${\mathcal N}\ldk\cdots\rdk$ denotes the normal ordering of particle and hole operators 
defined by (\ref{newbase1}). 
The HF single-particle energy is determined from the self-consistent equation 
\cite{Miyakawa_2008,Sogo_2009}
\beq
\epsilon_{\bk}&=& 
\epsilon_{1\bk} 
+\frac{1}{2}
\sum_{\bq} 
V_{dd}(\bq-\bk) f_\bq
\label{HFeq1}
\eeq
where $f_\bq=\theta\lk \epsilon_F-\epsilon_{\bq} \rk$ 
the Fermi-Dirac distribution function at zero temperature. 
In the case of the dipole moments polarized along the $z$ axis, 
the HF single-particle energy is well-described using two parameters 
$\lambda$ and $\beta$ \cite{Ronen_2010,Nishimura_2021}: 
\beq
\epsilon_{\bk} 
&=&\epsilon_{0} + \lambda^2 \frac{\beta^{-1} \lk k_x^2+k_y^2\rk +\beta^2 k_z^2}{2m_1}. 
\label{HFsol1}
\eeq 
Here $\epsilon_{0}$ is the rest energy, 
the parameters $\lambda$ ($\lambda^2\ge1$) and $\beta$ ($0<\beta \le 1$) 
determine the strength of the effective mass and 
the anisotropy in momentum space, rspectively. 
For instance, 
the perturbation theory gives 
the explicit formula $\beta=1-\frac{2m_1\bd^2k_F}{9\pi}$ \cite{Ronen_2010}, 
and it can be evaluated in the variation method \cite{Sogo_2009}. 
The parameter $\lambda^2$ takes values of the order of unity:
as an example, 
$\lambda^2 \simeq 1.00037$ 
for Dy atoms with a peak density of $4\times 10^{13} {\rm cm}^{-3}$\cite{MNY_2023}.
Using these parameters, the Fermi energy becomes 
$\epsilon_{F}=\epsilon_{0}+\lambda^2 k_F^2/2m_1$,
where $k_F=\lk 6\pi^2n_f\rk^{1/3}$  with $n_f$ the density of dipolar fermions.

\section{induced interaction between two impurity atoms} 
Now we consider two impurity atoms immersed in the medium of the degenerate dipolar Fermi gas.  
The impurity atoms interact with each other 
via the induced interaction in the medium even if no direct interaction exists in vacuum. 
Since the interaction between impurity and medium atoms is of the density-density type; 
such induced interaction should be mediated by the density fluctuation of the medium, 
which is described in terms of the density-density correlation function \cite{FW_2003} 
defined in the frequency $\omega$ and the momentum $\bq$ space by 
\beq
\Pi(\bq,\omega) &=& -i \int \rd t \, \rd^3\bx \, e^{-i\omega t + i \bx\cdot \bq} 
\left\langle {\mathcal T}\ldk \hat{n}(x) \hat{n}(0) \rdk \right\rangle, 
\nn 
&=&
-i
\int \rd t e^{-i\omega t}  \nn
&&\times
\left\langle
{\mathcal T}\ldk 
c_{1\bk}^\dagger(t)  c_{1\bk+\bq}(t)
c_{1\bk'+\bq}^\dagger(0) c_{1\bk'}(0)
\rdk 
\right\rangle. 
\eeq
In this study we approximate the correlation function to the single-loop order, 
and denote its static limit $\omega\rightarrow 0$ by $\Pi_{0}$: 
\beq
\Pi_{0}(|\tilde{\bq}_\beta|) &=& 
\sum_\bk \frac{f_\bk - f_{\bk+\bq}}{\epsilon_{\bk} - \epsilon_{\bk+\bq}}
\nn
 &=& 
-\frac{m_1 k_F}{4\pi^2\lambda^2} 
\lk 1 + \frac{4-|\tilde{\bq}_\beta|^2}{4 |\tilde{\bq}_\beta|}\ln\left| 
\frac{|\tilde{\bq}_\beta|+2}{|\tilde{\bq}_\beta|-2}\right|\rk 
\label{correl1}
\eeq
where we indicate the dependence on parameter $\beta$ explicitly 
through the dimensionless momentum $\tilde{\bq}_\beta$ defined by 
\beq
\tilde{\bq}_\beta=k_F^{-1} \lk \beta^{-1/2}q_x, \beta^{-1/2}q_y, \beta q_z\rk.  
\label{qtildeb1}
\eeq
For the detailed derivation of the correlation function (\ref{correl1}),
see appendix~\ref{app:A}. 

The inverse Fourier transform of the correlation function at the static limit gives 
the two-body potential of the induced interaction in real space:
\beq
V(|\tilde{\br}_\beta|) &=& g^2\sum_\bq \, e^{i \bx\cdot \bq}\, \Pi_0(|\tilde{\bq}_\beta|) 
\nn 
&=&
g^2\frac{m_1 k_F^4}{16\pi^3\lambda^2} 
\lk
\frac{2\cos2|\tilde{\br}_\beta|}{|\tilde{\br}_\beta|^3} -\frac{\sin2|\tilde{\br}_\beta|}{|\tilde{\br}_\beta|^4}
\rk, 
\label{rkky0}
\eeq
where we have introduced 
a $\beta$-dependent dimensionless coordinate $\tilde{\br}_\beta$:  
\beq
\tilde{\br}_\beta&=&\lk \beta^{1/2}x, \beta^{1/2}y, \beta^{-1} z\rk k_F. 
\label{coordinate1}
\eeq
For the derivation in detail, see appendix~\ref{app:B}. 
The resultant potential (\ref{rkky0}) shows the RKKY type, 
but the spacial anisotropy exists through the space coordinates in $\tilde{\br}_\beta$, 
which is originated in the the momentum-space anisotropy in the correlation function. 
The polarization direction of the dipole moment can be taken arbitrarily; 
then one can replace $\tilde{\br}_\beta$ in (\ref{coordinate1}) by $\tilde{\bx}_\beta$:
\beq 
\tilde{\bx}_\beta
&=& k_F \beta^{1/2}\lk \bx - \bx\cdot \hat{\bm{d}}\hat{\bm{d}}\rk 
+ k_F \beta^{-1}  \bx\cdot \hat{\bm{d}}\hat{\bm{d}} 
\label{coordinate2}
\eeq
where $\hat{\bm d}=\bd/|\bd|$. 
Since the anisotropic RKKY potential is a function of $|\tilde{\bx}_\beta|$, 
it always has the rotational symmetry around the dipole moment. 
In the rest of this paper, we mainly take the polarization direction along $z$-axis 
except in Sec.~V-A where the arbitrary directions are for scattering problem.  

\section{Two-body bound states}
We investigate the two-body problem for impurity atoms interacting with the anisotropic RKKY potential (\ref{rkky0}), 
in which impurity atoms are treated quantum mechanically. 
As a rigorous treatment for the in-medium two-body problem, 
one can employ, for instance, the in-medium $T$ matrix approach \cite{Kashimura_2012,Tajima_2018}, 
which incorporates the self-energy effects
such as single-particle residue, effective mass, and decay width, 
or conventional Br\"{u}ckner's $G$-matrix theory 
for the effective interaction in fermionic medium \cite{RS_2004}. 
In the present treatment, on the other hand, we assume the case where 
the quasiparticle picture is well established
for individual dressed impurity atom (polaron); 
the self-energy effects from loop corrections 
are small in comparison with mean-field effects,  
the quasiparticle residue is not far from unity 
and the inverse decay width is much smaller than the polaron rest energy.  
According to the previous work \cite{Nishimura_2021}, 
the polaron dispersion relation in the dipolar Fermi gas is given in the form  
\beq
\mathcal{E}_\bp &=& \mathcal{E}_0 + \frac{p_x^2+p_y^2}{2m_t}+ \frac{p_z^2}{2m_z} 
+\mathcal{O}\lk p^4\rk, 
\eeq
where $\mathcal{E}_0$ is the mean-field rest energy due to the impurity-medium interaction, 
and $m_{t,z}$ represent anisotropic effective masses reflecting the Fermi surface deformation. 
In this study we ignore the anisotropic effect on the effective mass,
$m_{t}=m_{z}=m_{2}$, 
since the corrections for the mass-difference can be estimated to be about $3 \%$ or less 
\cite{Nishimura_2021}. 
Also, it is to be noted that 
since the introduction of single impurity atom into the medium 
costs $\mathcal{E}_0$ which is negative  (positive) for $a_{12}<0$ ($a_{12}>0$), 
the binding energy to be evaluated in this paper 
should be measured from the two-impurity atom threshold $2\mathcal{E}_0$.

We first evaluate the condition of the bound-state formation in parameter space, 
by solving the Schr\"{o}dinger equation for the relative coordinates between two impurity atoms, 
\begin{widetext}
\beq
E \psi(\bx)&=& \ldk -\frac{k_F^2}{2m_{22}\beta^2}\partial_z^2 
-\beta \frac{k_F^2}{2m_{22}}\lk \frac{1}{r^2}\partial_\theta^2 + \frac{1}{r} \partial_r r \partial_r \rk + 
V(\sqrt{r^2+z^2}) \rdk \psi(\bx) 
\eeq
\end{widetext}
where $m_{22}=m_2/2$ is the reduced mass.
We have employed 
the cylindrical coordinates and scaled the variables as 
\beq
\bx=k_F^{-1} \lk \beta^{-1/2} r\cos\theta, \beta^{-1/2}r\sin\theta, \beta z\rk, 
\eeq
so that the anisotropic parameter $\beta$ appears only in the kinetic term 
and the space coordinates $(r,z)$ become dimensionless. 
This manipulation reduces the numerical costs significantly. 

Substituting the partial-wave expansion of the wave function for the angular momentum of $z$ component ${l_z}$: 
\beq
\psi(\br) = \sum_{{l_z}=0,\pm1,\pm2,\cdots} \frac{1}{\sqrt{2\pi}} e^{il_z\theta} \psi_{l_z}(r,z), 
\eeq 
into the Schr\"{o}dinger equation,
we obtain the equation for each $l_z$ component: 
\begin{equation}
0= \ldk \frac{1}{\beta^2}\partial_z^2 
+\beta\frac{1}{r}\partial_r r \partial_r 
+\beta\frac{{l_z}^2}{r^2} - v(\sqrt{r^2+z^2}) +\varepsilon \rdk \psi_{l_z}, 
\label{scheq1} 
\end{equation}
where 
$E=\frac{k_F^2}{2m_{22}}\varepsilon$ and $V(r) =\frac{k_F^2}{2m_{22}} v(r)$. 
We have used the dimensionless potential:
\beq
v(r)=G\lk \frac{2\cos2r}{r^3} -\frac{\sin2r}{r^4} \rk,
\label{dimensionlessrkky1}
\eeq 
where the dimensionless coupling constant $G$ is defined by  
\beq
G \equiv \frac{2m_{22}}{k_F^2} g^2\frac{m_1 k_F^4}{16\pi^3\lambda^2}.
\label{dimensionlessG1}
\eeq

\subsection{Finite-range boundary conditions}
To solve the eigenvalue problem of Eq.~(\ref{scheq1}) numerically, 
we impose finite boundary conditions; the system is  
periodic in $z$ direction with interval $-L/2\le z \le L/2$, 
and confined in radial direction within $0\le r\le R$. 
As the complete orthonormal systems for wave-function expansion,
we use
\begin{widetext}  
\beq
u_n(z) &:=& 
\ltk 
\begin{array}{c}
\sqrt{\frac{2-\delta_{n,0}}{L}} \cos\frac{2\pi n z}{L}, \ {\rm for} \  l_z=0,2,4,\cdots \\
\sqrt{\frac{2}{L}} \sin\frac{2\pi n z}{L}, \ {\rm for} \  l_z=1,3,5,\cdots 
\end{array} 
\right., 
\quad 
\int_{-L/2}^{L/2} {\rm d}z \, u_m^*(z) u_n(z) = \delta_{mn}, 
\label{func1}
\\
J_{{l_z};i}(r) &:=& \frac{\sqrt{2}}{R J_{{l_z}+1}(s_i)} J_{l_z}(s_i r/R), \quad 
 \int_{0}^{R} {\rm d}r r\,  J_{{l_z};i}(r)  J_{{l_z};j}(r) = \delta_{ij}, 
 \label{func2}
\eeq
where $n=0,1,2,\dots$, and 
$s_i$ is the zero's of the Bessel function, $J_{l_z}(s_i)=0$ 
($s_{1}< s_{2} < s_{3} < \cdots$).
Then, the wave function is expanded by 
\begin{equation}
\psi_{l_z}(r,z) = \sum_{n=0,1,2,\cdots} \sum_{i=1,2,3,\cdots} f_{l_z}(n;i)  u_n(z) J_{{l_z};i}(r). 
\label{wf1}
\end{equation}
The eigenvalue problem reduces to the matrix form: 
\beq
0&=& 
\sum_{n=0,\pm1,\pm2,\cdots} \sum_{j=1,2,3,\cdots} \, 
\ldk 
\lk\frac{1}{\beta^2}\lk\frac{2\pi n}{L}\rk^2+ \beta\lk\frac{s_i}{R}\rk^2 
-\varepsilon \rk \delta_{mn}\delta_{ij}
 +  v_{mi;nj}
\rdk \, f_{l_z}(n;j) 
\label{matrixeq1}
\eeq
where the matrix elements for the RKKY potential are defined by 
\beq
v_{mi;nj} &\equiv& 
\int_{-L/2}^{L/2} {\rm d}z \,  
\int_0^R  {\rm d}r \, r\, u_m^*(z) \, J_{{l_z};i}(r)\, 
v(\sqrt{r^2+z^2}) \, u_n(z) \, J_{{l_z};j}(r). 
\eeq
The summation in the matrix equation (\ref{matrixeq1}) is taken within  
the truncated numbers $n_{\rm max}$ and $i_{\rm max}$:
$|m|,|n|\le n_{\rm max}$ and $i, j\le i_{\rm max}$. 
\end{widetext}

\subsection{Numerical results for bound states} 
\begin{figure}[h]
  \begin{center}
    \begin{tabular}{l}
 \resizebox{83mm}{!}{\includegraphics{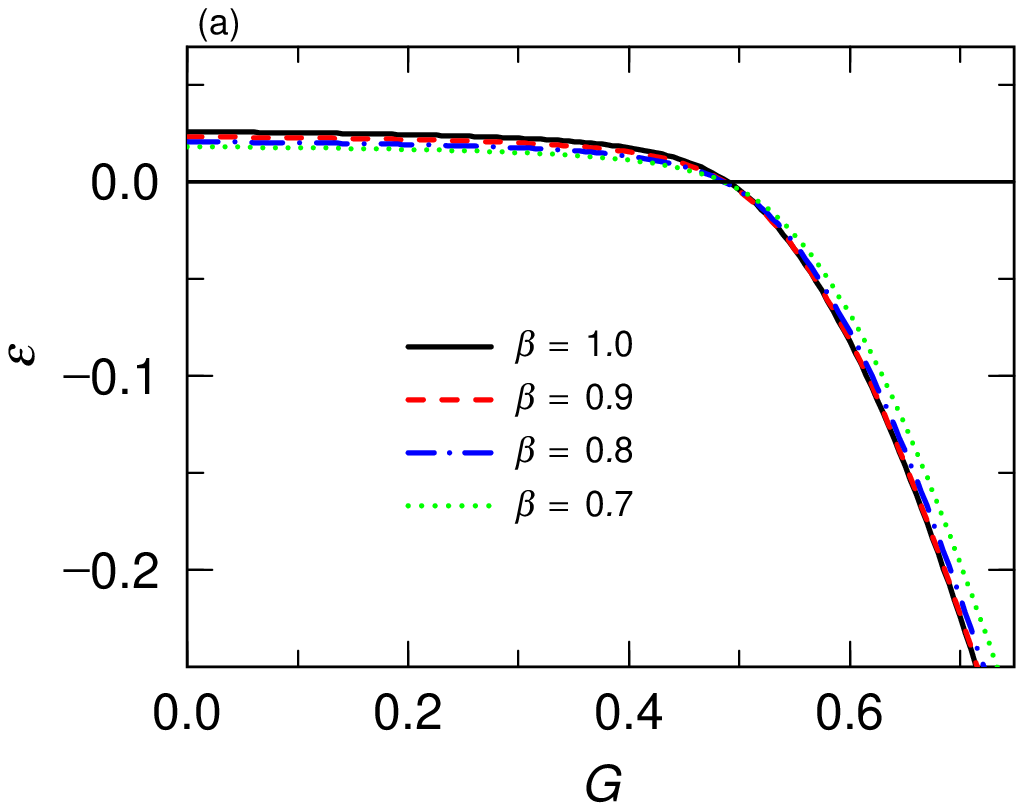}} \\
 \resizebox{83mm}{!}{\includegraphics{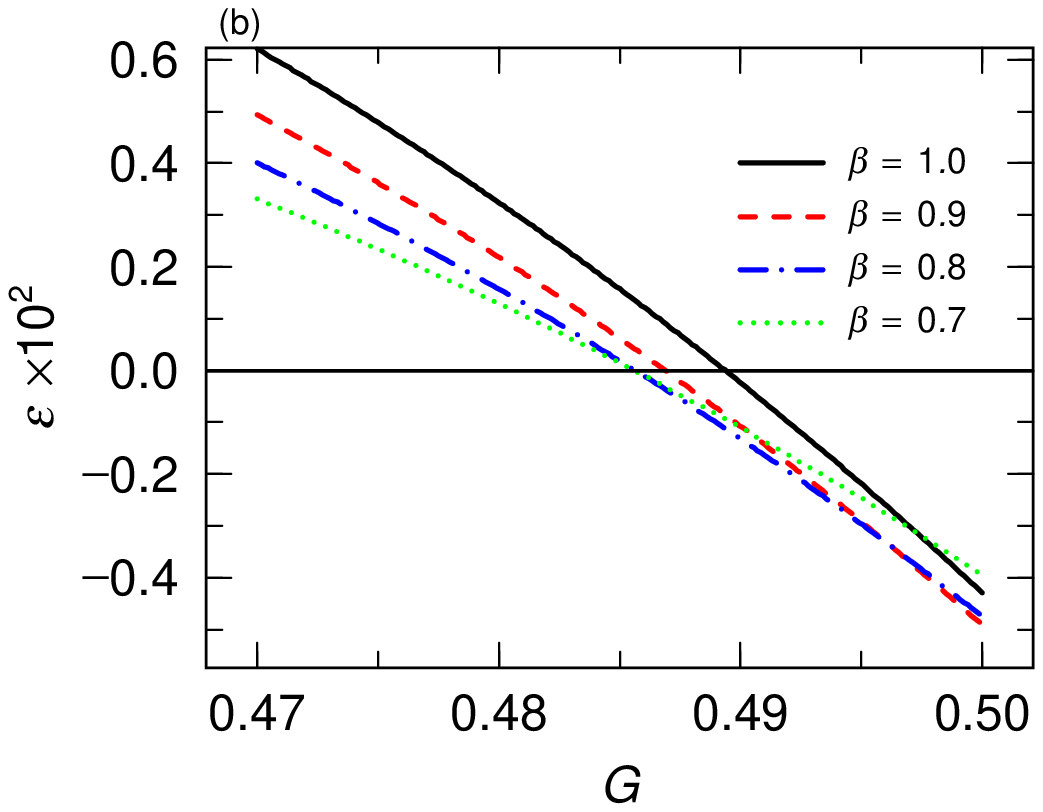}} 
    \end{tabular}
    \caption{(a) the lowest energy eigenvalue (the first bound state for $l_z=0$) 
    as a function of coupling constant $G$, 
    determined from Eq.~(\ref{matrixeq1}) 
    in the case that $n_{\rm max}=i_{\rm max}=30$, $L=2R=30 k_F^{-1}$.  
    Different lines correspond to $\beta=1.0, 0.9, 0.8, 0.7$, respectively. 
    (b) blow-up of critical region. }
    \label{energyvsG1}
  \end{center}
\end{figure}

In Fig.~\ref{energyvsG1}
we show the value of the lowest energy for $l_z=0$ as a function of 
the dimensionless coupling strength $G$ in (\ref{dimensionlessG1}) 
for various values of $\beta$, 
in which we have taken the size of the system and the dimension of matrix large enough 
for convergence.  
The result does not depend on $\beta$ significantly, 
which can be understood from the fact that the $\beta$-dependence 
of the kinetic-energy in (\ref{matrixeq1}) is saturated around $\beta=1$.  

From the numerical results, we can also determine 
the critical value of the coupling constant, $G_{\rm crit}$, 
where the first bound state just emerges, i.e., $\varepsilon = 0$ in the finite system. 
In order to estimate $G_{\rm crit}$ in the spatially uniform system, 
we have employed an extrapolation procedure 
to find the critical value at $R, L \rightarrow \infty$.
The results are summarized in Table~\ref{critG1}. 
For the detail of the extrapolation procedure, see the appendix~\ref{app:C}. 
\begin{table}[h]
\begin{center}
\begin{tabular}{|c|c|c|c|c|} \hline
    & $\beta = 1.0$ & \hspace{0.22cm} $0.9$ \hspace{0.22cm}  &  \hspace{0.22cm} $0.8$ \hspace{0.22cm}  &  \hspace{0.22cm}  $0.7$  \hspace{0.22cm}  \\ \hline
$G_{\rm crit}$ & $0.488$ & $0.485$ & $0.4843$ &  $0.4847$    \\ \hline
\end{tabular}
\end{center}
\caption{Estimations of critical coupling constant for the first bound state of $l_z=0$ in the uniform system 
for some different values of $\beta$.} 
\label{critG1}
\end{table}
The result implies that the critical value decreases much dully and seems saturated around $\beta=0.7 \sim 0.8$. 

To check the accuracy in the present numerical calculation, 
we have examined the case of the spherical symmetry, i.e., $\beta=1$, 
by solving the same problem in the polar coordinates with the dimension of integral reduced, 
and found that the critical coupling is estimated to be $G_{\rm crit}=0.501$, 
which differs by a few $\%$ from $G_{\rm crit}=0.488$ in the cylindrical coordinates. 
The details of the calculation in the spherical case is presented in appendix~\ref{app:D}. 
In Fig.~\ref{rkkypotvsrz2}, we show the RKKY potential (\ref{dimensionlessrkky1}) 
and the wave functions of the first bound state in the spherical case ($\beta=1$), obtained 
from both cylindrical and polar coordinates systems, and find that the results seem to be consistent as a whole. 

\begin{figure}[hbtp]
    \begin{tabular}{l}
 \resizebox{83mm}{!}{\includegraphics{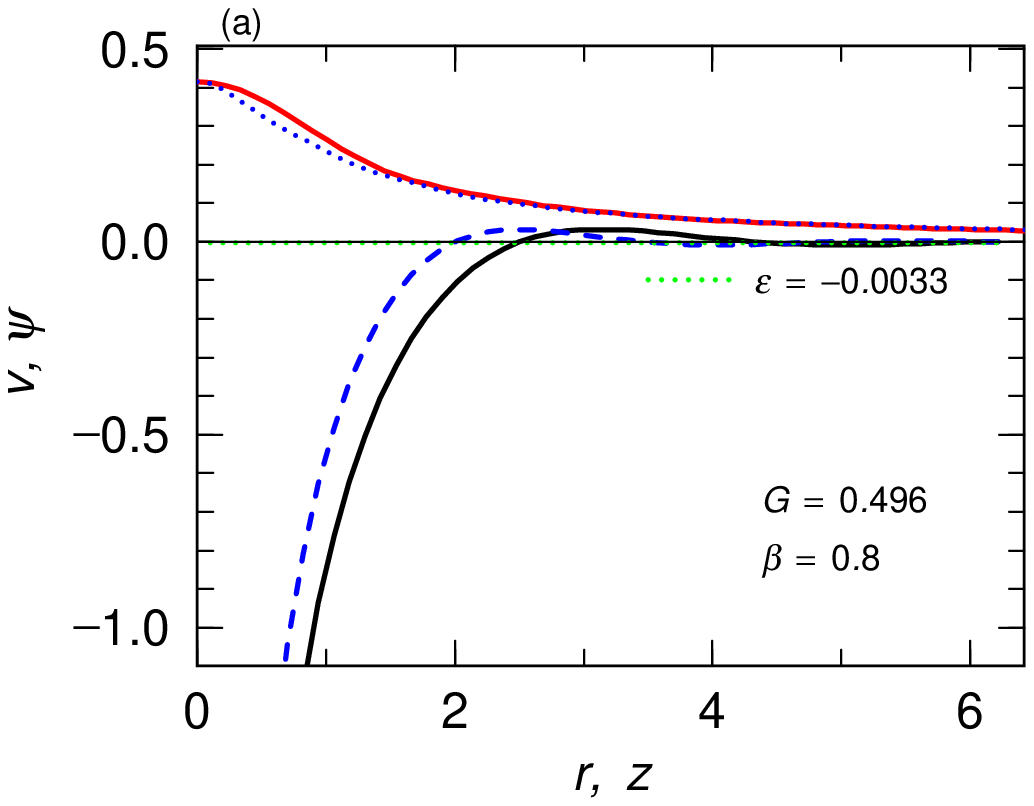}} \\
 \resizebox{83mm}{!}{\includegraphics{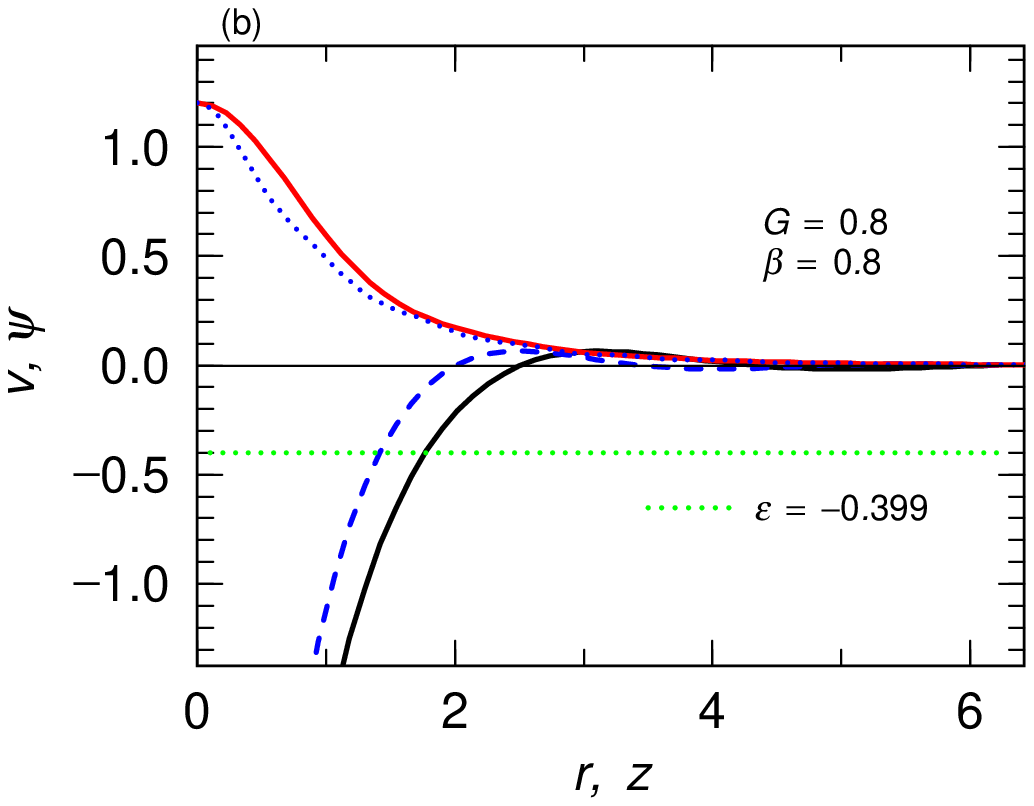}} \\ 
 \resizebox{83mm}{!}{\includegraphics{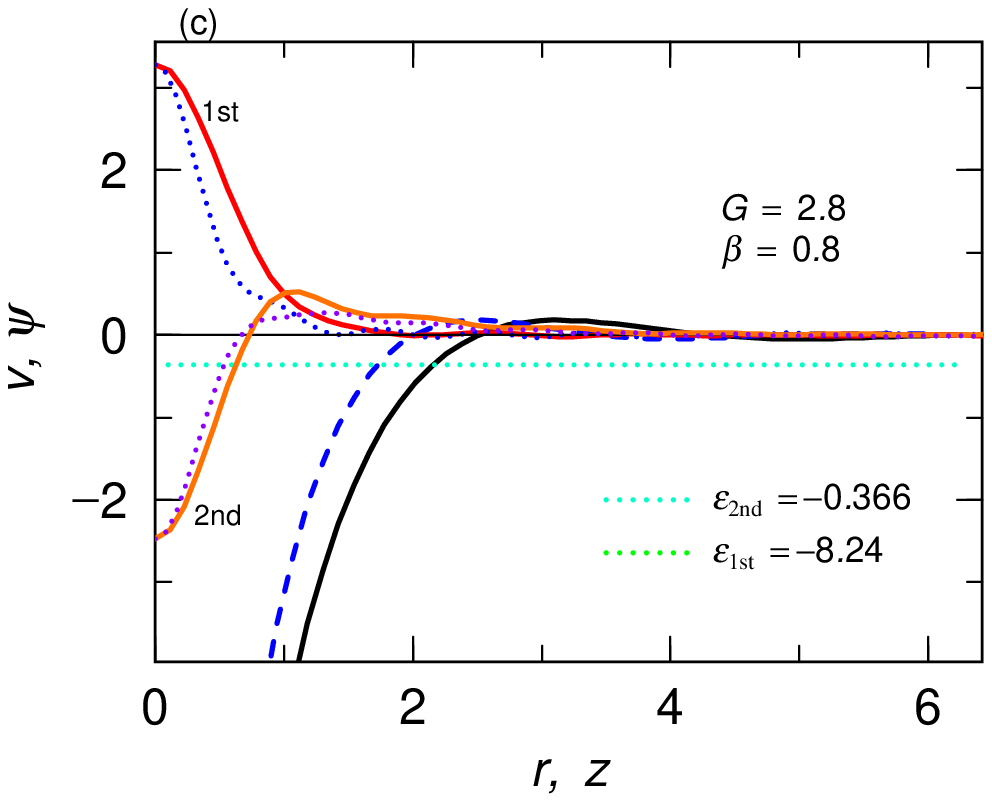}} 
    \end{tabular}
    \caption{Numerical results of bound states for $\beta=0.8$, $n_{\rm max}=i_{\rm max}=30$, and $L=2R=30$.  
       Figures (a), (b), and (c) correspond to $G=0.496$, $0.8$, and $2.8$, respectively. 
     Solid and dashed lines rising from negative side represent the dimensionless potential $v(\beta^{1/2}r)$ as functions of $r$ ($z=0$), and $v(\beta^{-1}z)$ as functions of $z$ ($r=0$) 
    in (\ref{dimensionlessrkky1}), respectively. 
     Horizontal dotted lines represent the level of binding energies, and 
    the corresponding wave functions (\ref{wf1})  in $r$ and $z$ directions, i.e., 
    $\psi(r,0)$(solid lines) and $\psi(0,z)$(dotted lines), are shown together. 
    }
    \label{rkkypotvsrz1}
\end{figure}

Finally in Fig.~\ref{rkkypotvsrz1} 
we show the potential shape and the bound-state wave functions 
together with the value of corresponding binding energy in the case of $\beta=0.8$ 
for different strengths of the coupling constant. 
We can observe in Fig.~\ref{rkkypotvsrz1}(a), (b) 
that, while at $G=0.496$ (just above the critical value)
the value of binding energy almost overlaps the modulating part of the RKKY potential around zero 
and the wave functions show long tails, 
they move deeper inside the potential at $G=0.8$ to become well-stabilized. 
At a larger coupling constant, for instance, at $G=2.8$,  
the second bound state emerges, whose wave functions are shown 
in Fig.~\ref{rkkypotvsrz1}(c) 
together with the first one.  
\begin{figure}[hbtp]
 \resizebox{85mm}{!}{\includegraphics{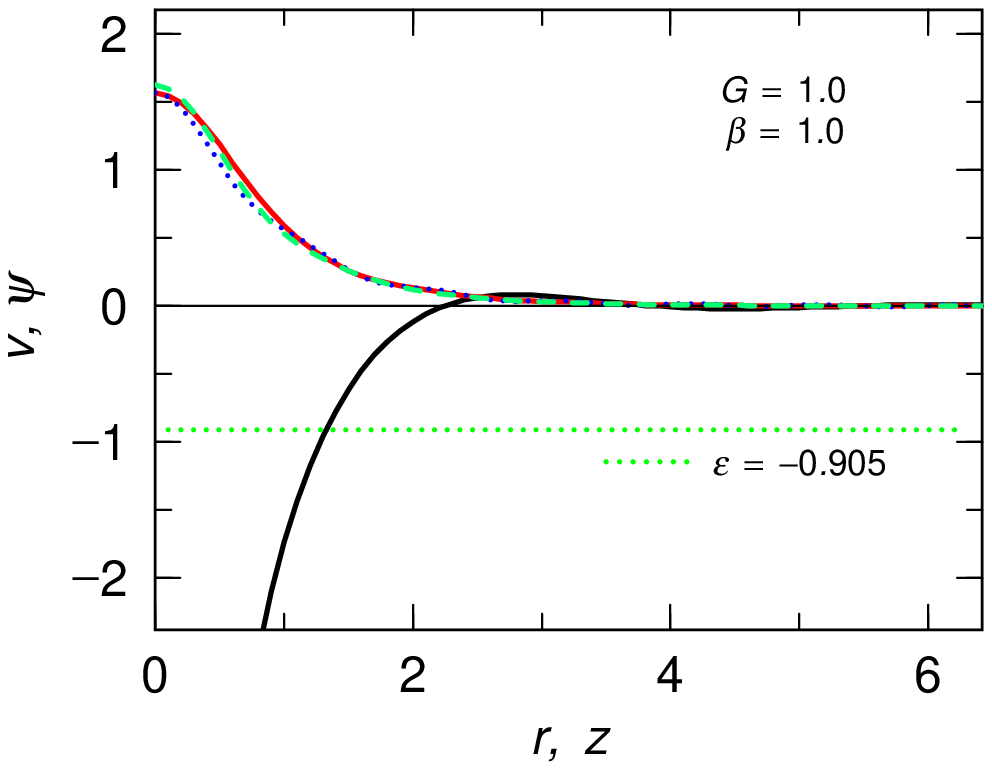}} 
    \caption{
    Wave functions of bound state in the spherically symmetric case $\beta=1$ at $G=1.0$. 
    The dimensionless potential $v(r)$ is presented by the modulated solid line.  
    The wave functions (\ref{wf1}) in $r$ and $z$ directions 
    are presented by solid and dotted lines, respectively. 
    We also show the wave function as a function of $r$ in the polar coordinates by dashed line, 
    which is made from spherical Bessel functions and scaled by some factor for comparison. 
    }
    \label{rkkypotvsrz2}
\end{figure}


\section{Two-body scattering}
A salient future of the present system is spatial anisotropy 
of the induced interaction potential; 
accordingly, the scattering process becomes anisotropic with respect to the direction of dipole polarization $\bd$. 
In the present section, 
we figure out explicitly the directional-dependence of the two-body scattering amplitude for impurity atoms 
on the angles among $\bd$, initial and final momenta $\bk$ and $\bk'$, 
and further make the partial-wave analysis 
to observe transitions between different angular-momentum states in the course of scatterings. 
It should be noted again that in the present analysis 
we ignore the self-energy effects on the single impurity atom, 
assuming especially that the life-time of the impurity quasiparticle (polaron) is 
much longer than the time-scale of scattering processes, 
which is realized at low temperatures below the Fermi energy. 

\subsection{Scattering amplitude}
We evaluate the scattering amplitude in the Born approximation. 
In the description of the scattering problem for the anisotropic potential, 
we take the $z$-axis of the impurity's relative coordinates 
to the direction of initial momentum $\bk$; 
those of $\bd$ and $\bk'$ are taken arbitrary.  
Since only the angle between $\bd$ and $\bk$ is relevant,  
the parametrization of $\bd$, $\bk$ and $\bk'$ are taken in the spherical coordinates as  
\begin{equation}
\hat{\bd} = 
\lk 
\begin{array}{c}
\sin\alpha\\
0\\
\cos\alpha
\end{array}
\rk, 
\ 
\bk= 
k
\lk 
\begin{array}{c}
0 \\
0\\
1
\end{array}
\rk,  
\ 
\bk' = 
k
\lk 
\begin{array}{c}
\sin\theta\cos\phi \\
\sin\theta\sin\phi\\
\cos\theta
\end{array}
\rk, 
\label{parameter1}
\end{equation}
where $k=|\bk'|=|\bk|$. 

In general the $T$ matrix for two-body scattering in the center of mass frame 
is determined from the Lippmann-Schwinger equation: 
\beq
\hat{T} &=& \hat{V} + \hat{V} \frac{1}{E-\hat{H}_0 +i\eta} \hat{T}.
\eeq
In momentum-space representation, 
it becomes 
\beq
\nn
T_{\bk \bk'} &=&  V_{\bk \bk'} 
+ \sum_\bq V_{\bk \bq} \frac{1}{E-\bq^2/2m_{22} +i\eta} T_{\bq \bk'} 
\eeq
where $2m_{22} E= \bk^2={\bk'}^2$, 
$T_{\bk \bk'} = \langle \bk | \hat{T} |\bk' \rangle$ 
and $V_{\bk \bk'} = \langle \bk | \hat{V} |\bk' \rangle$. 
The scattering amplitude $f_{\bk,\bk'}$ is represented by
\beq
f_{\bk,\bk'} &=& -\frac{m_{22}}{2\pi} T_{\bk \bk'}.  
\eeq
In the Born approximation, 
we take the leading-order term in the iterative expansion for the $T$ matrix : 
\beq
T_{\bk \bk'} &=& \langle \bk | \hat{V} |\bk' \rangle 
\nn
&=& \sum_{\bx} 
V(|\tilde{\bx}_\beta|) e^{i \bx \cdot \lk \bk'-\bk\rk} 
\nn
&=& 
g^2\sum_{\bx,\bq} \, e^{i \tilde{\bx}_\beta \cdot \tilde{\bq}_1}\, \Pi_0\lk |\tilde{\bq}_1|\rk 
e^{i \bx \cdot \lk \bk'-\bk\rk} 
\nn
&=& 
g^2 
\, \Pi_0(\tilde{k}_\beta)\, 
\label{scatteringamp5}
\eeq
where  $\tilde{\bq}_1=\left.\tilde{\bq}_{\beta}\right|_{\beta=1}$ in (\ref{qtildeb1}), and 
we have employed the potential $V(|\tilde{\bx}_\beta|)$ (\ref{rkky0}) 
together with (\ref{coordinate2}).
The magnitude of the scaled momentum $\tilde{k}_\beta$ is defined by 
\begin{widetext}
\beq
\tilde{k}_\beta
&=& 
k_F^{-1}
\sqrt{
 \beta^{-1} 
\ldk -k_x' \cos\alpha + \lk k_z'-k\rk \sin\alpha\rdk^2 
+
\beta^{-1} k_y'^2
+
\beta^2 
\ldk k_x' \sin\alpha + \lk k_z'-k\rk \cos\alpha\rdk^2
} 
\nn
&=& 
k_F^{-1}k
\ltk 
 \beta^{-1} 
\ldk -\sin\theta\cos\phi \cos\alpha + \lk \cos\theta-1\rk \sin\alpha\rdk^2 
\right.
\nn
&&\left.
\qquad \quad +
\beta^{-1} \lk\sin\theta\sin\phi\rk^2
+
\beta^2 
\ldk \sin\theta\cos\phi \sin\alpha 
+ \lk \cos\theta-1\rk \cos\alpha\rdk^2
\rtk^{1/2}. 
\eeq
In the derivation of (\ref{scatteringamp5}) we have also used the summation formula:
\beq
\sum_{\bx} e^{i\tilde{\bx}_\beta \cdot \tilde{\bq}_1 + i\bx\cdot\lk \bk'-\bk\rk} 
&=& \lk 2\pi\rk^3 \delta^{(3)}\ldk 
k_F \beta^{1/2}\bq
+ k_F \lk \beta^{-1} - \beta^{1/2} \rk \hat{\bm{d}}\cdot \bq \hat{\bm{d}}
+ \bk'-\bk \rdk.  
\eeq

In Fig.~\ref{fkk1}, 
we show the angle dependence of scattering amplitude at a low energy 
for various values of $\alpha$, the angle between $\bd$ and $\bk$. 
The results show that the forward scattering ($\theta \simeq 0$) dominates as a whole, 
which is usually expected for non-singular interaction potentials. 
At $\alpha=0$ the rotational symmetry leads to independence of 
the scattering amplitude from the azimuthal angle $\phi$ 
in the similar manner as the case of spherically symmetric potentials. 
In cases of $\alpha\neq0$, on the other hand, 
the scattering amplitude gradually depends on $\phi$ when $\theta\neq0$ 
and develops a maximal peak 
at $\phi=\pi$ ($\phi=0$) for $0< \alpha \le \pi/2$ ($\pi/2\le \alpha \le \pi$). 
This result implies that a dilute gas of impurity atoms is expected to exhibit 
anisotropic properties in transports \cite{Aikawa_2014c,Wang_2021} 
and in propagation of collective excitations \cite{Chan_2010,Ronen_2010,Sieberer_2011,MNY_2023}.   
\begin{figure}[h]
  \begin{center}
    \begin{tabular}{lll}
 \resizebox{57mm}{!}{\includegraphics{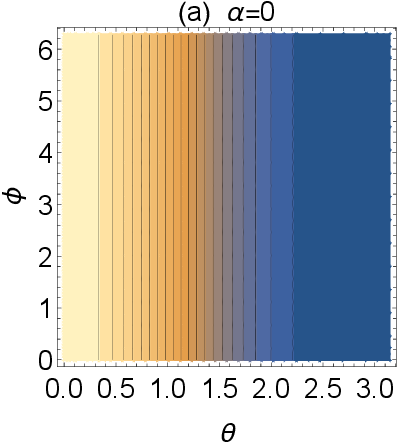}} & \  
 \resizebox{57mm}{!}{\includegraphics{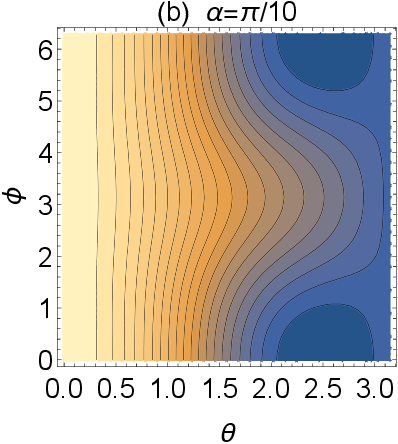}} & \  
 \resizebox{57mm}{!}{\includegraphics{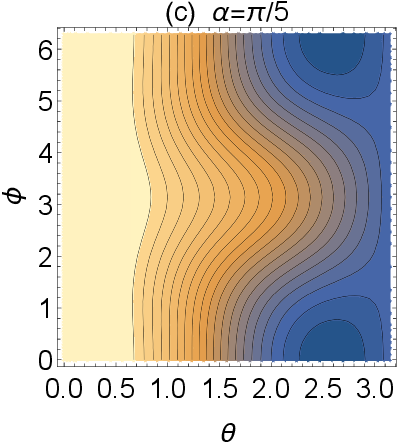}} \\
 \resizebox{57mm}{!}{\includegraphics{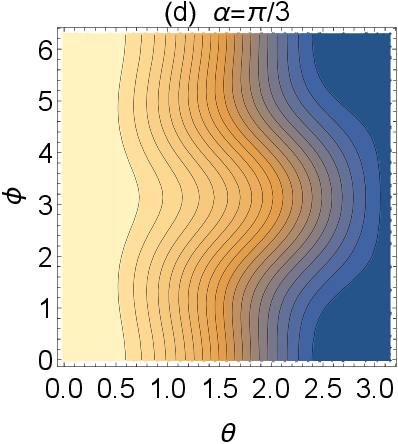}} & \ 
 \resizebox{57mm}{!}{\includegraphics{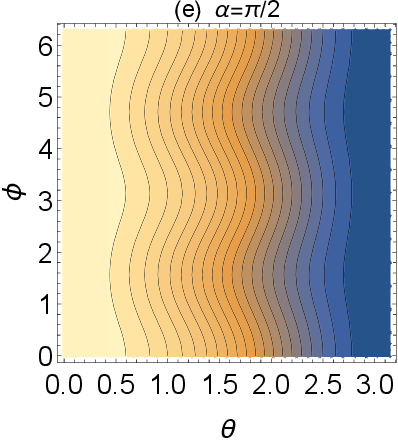}} & \ 
 \resizebox{57mm}{!}{\includegraphics{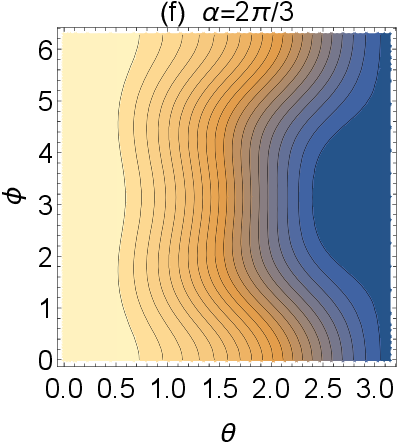}} 
    \end{tabular}
    \caption{Contour plots of scattering amplitude $f_{\bk,\bk'}$ for $k=0.2k_F$ and $\beta=0.8$ in $\theta$-$\phi$ plane. 
    From left up to right down, $\alpha$, the angle between $\bk$ and $\bd$, varies from $0$ to $2\pi/3$. 
    The contours in smaller values of $\theta$, i.e., the forward scattering region, take larger values. 
  Numerical values of contours typically ranges from $1.99$ (highest) to $1.97$ (lowest) 
  in the unit of $m_{22}g^2/2\pi$. }
    \label{fkk1}
  \end{center}
\end{figure}
\end{widetext}

In relation to such dilute gas of impurity atoms at low temperatures, 
we can incidentally derive the $s$-wave scattering length 
$a_s$ for the medium induced RKKY potential. 
In the low energy limit the correlation function becomes 
$\lim_{q\rightarrow 0} \Pi_{0}(q) = -2\frac{m_1 k_F}{4\pi^2\lambda^2}$, 
thus $a_s$ can be read off as 
\beq
\lim_{k\rightarrow 0} 
f_{\bk,\bk'} &=& 
\lk -\frac{m_{22}}{2\pi} \rk g^2 
\lk -2\frac{m_1 k_F}{4\pi^2\lambda^2}\rk \nn &=&
g^2\frac{m_{22} m_1 k_F}{4\pi^3 \lambda^2} \equiv -a_s. 
\eeq
The scattering length is negative as expected from the attractive nature of RKKY potential 
attributed to the fermionic single-loop contribution, 
and this result should be compatible with the validity condition of the Born approximation 
at the low energy limit: 
\beq
\frac{2}{\pi} \gg |a_s| k_F. 
\eeq
See appendix~\ref{app:E} for the derivation of this condition. 

\subsection{Partial-wave analysis}
In the presence of the anisotropic RKKY potential for $\beta\neq1$ , 
the spherical symmetry reduces to uniaxial one around the dipole moment. 
Therefore, the angular momentum is no longer conserved in the scattering processes. 
To see this quantitatively, we make the partial-wave analysis using 
the expansion of plane wave in terms of spherical harmonics 
\beq
e^{i\bk\cdot \bx} &=& \sum_{l=0}^\infty i^l (2l+1) j_l\lk kr\rk P_l\lk \frac{\bk\cdot \bx}{kr} \rk
\nn
&=& 4\pi \sum_{l=0}^\infty \sum_{m=-l}^l  i^l j_l\lk kr\rk {Y^*}_l^m\lk\Omega_p\rk Y_l^{m}\lk\Omega_x\rk
\eeq
where 
$\Omega_x$ and $\Omega_k$ represent the solid angles of $\bx$ and $\bk$, respectively. 
Then 
we can express the $T$ matrix in the Born approximation as 
\begin{widetext}
\beq
T_{\bk \bk'} &=& \langle \bk | \hat{V} |\bk' \rangle 
\nn
&=& \sum_{\bx} 
V(\bx) e^{i \bx \cdot \lk \bk'-\bk\rk} 
\nn
&=&
(4\pi)^2
\sum_{\bx} 
\sum_{l=0}^\infty \sum_{m=-l}^l 
 i^{l} j_{l}\lk kr\rk {Y}_{l}^{m}\lk\Omega_{k}\rk {Y^*}_{l}^{m}\lk\Omega_x\rk
V(\bx)
\sum_{l'=0}^\infty \sum_{m'=-l'}^{l'} 
 i^{l'} j_{l'}\lk kr\rk Y_{l'}^{m'}\lk\Omega_x\rk {Y^*}_{l'}^{m'}\lk\Omega_{k'}\rk 
\nn
&=&
\sum_{l,l'} \sum_{m,m'} 
 {Y}_{l}^{m}\lk\Omega_{k}\rk
 V_{lm; l'm'}(k) 
 {Y^*}_{l'}^{m'}\lk\Omega_{k'}\rk 
\eeq
where we have defined the matrix element by 
\beq
V_{lm; l'm'}(k) &=& (4\pi)^2  i^{l+l'}
\int_0^\infty \rd r r^2\, \int \rd\Omega_x
{Y^*}_{l}^{m}\lk\Omega_x\rk
j_{l}\lk kr\rk 
V(\bx)
j_{l'}\lk kr\rk Y_{l'}^{m'}\lk\Omega_x\rk.  
\label{me1}
\eeq
\end{widetext}
Here 
it should be noted that in the present analysis 
we have chosen the direction of $\bd$ parallel to $z$ axis; 
accordingly $V(\bx)$ used in the matrix element (\ref{me1}) is given by (\ref{rkky0}).  
Since the potential is symmetric about the rotation around the $z$ axis, 
the matrix element becomes diagonal with respect the magnetic quantum numbers: 
$V_{lm; l'm'}(k)\propto \delta_{m,m'}$, 
while the angular momentum may change from $l$ to $l'$ in the course of scatterings. 
In TABLE~\ref{matrixelem1}, 
we present a sample of the matrix element $V_{lm;l'm'}$ 
in the unit of  the $s$-wave element $V_{00;00}$ 
for a low energy scattering. 

It is found that among the diagonal elements 
the $s$-wave contribution dominates at low energy, 
and also that because of the parity conservation 
transitions between different parity states, i.e., 
from odd(even) $l$ to even(odd) $l'$, are prohibited. 
Now we consider the particle statistics of impurity atoms. 
For instance, in the case that impurity atoms are single-component fermions, 
$p$-wave scattering should be dominant at low energies, and 
transitions only with odd numbers of $l, l'$ are allowed. 
This observation implies that a dilute gas of single-component impurity Fermi atoms 
exhibits $p$-wave superfluidity with a small fraction of $f$-wave pairs 
in the medium of degenerate dipolar Fermi gas at low temperatures. 
\begin{widetext}
\begin{center}
\begin{table}[h]
\begin{tabular}{|c|c|c|c|c|c|c|} \hline
\diagbox{$\quad l,m$}{$l',m'\quad$}
& 0, 0 & 1, 0 & 1, $\pm1$ &  2, 0 & 2, $\pm1$  &  2, $\pm2$   \\ \hline
0, 0  & 1    & $0$ & 0 &  $-3.1\times 10^{-4}$ & 0 & 0 \\ \hline
1, 0  & $0$ & $8.0 \times 10^{-4}$ & 0 & $0$ & 0 & 0 \\ \hline
1, $\pm1$ & 0 & 0 & $1.6 \times 10^{-3}$ & 0 & $0$& 0 \\ \hline
2, 0  & $-3.1\times 10^{-4}$ & $0$ & 0 & $-2.4\times 10^{-5}$ & 0 & 0 \\ \hline
2, $\pm1$  & 0 & 0 & $0$ & 0 & $-1.2\times 10^{-6}$  & 0 \\ \hline
2, $\pm2$  & 0 & 0 & 0 & 0 & 0  & $-3.6 \times 10^{-5}$ \\ \hline
\end{tabular}
\caption{Matrix element $V_{lm;l'm'}/V_{00;00}$ for 
$k=0.1 k_F$, $\beta=0.8$. $V_{00;00}=-0.3175 V_0$ with 
$V_0=g^2\frac{m_1 k_F^4}{4\pi^3\lambda^2}=|a_s|\frac{k_F^3}{m_{22}}$.} 
\label{matrixelem1}
\end{table}
\end{center}
\end{widetext}

\section{Experimental example} 
Here we estimate numerical values of parameters relevant to the two-body problem 
from a gaseous mixture of dipolar Fermi and non-dipolar Fermi atoms, which is experimentally accessible.  
In the experiment of a gaseous mixture of polarized dipolar ${}^{161}$Dy atoms 
and non-dipolar ${}^{40}$K atoms \cite{Ravensbergen_2018}, 
the number imbalance is set to be relatively large $n_{\rm Dy}/n_{\rm K} \sim 5$, and 
the temperature of majority gas of ${}^{161}$Dy is lowered to $T \sim 45\, {\rm nK}$, 
much smaller than Fermi energy $T k_B/E_F \sim 0.09$, that is to say, 
the Fermi degeneracy is achieved. 
Here $n_{\rm Dy}$($n_{\rm K}$) is a peak density of Dy(K) atoms in trap, 
and $n_{\rm Dy} \sim 10^{14} \, {\rm cm}^{-3}$ is left after evaporative cooling. 
From this experimental situation we estimate the parameters 
$G = \lk 1 + r_m^{-1} \rk \lk 1 + r_m \rk\lk a_{12}k_F\rk^2/4\pi\lambda^2$ 
and $a_s k_F = -2 G$ as follows:
\beq
 7.3\times10^{-4} \le & G & \le 4.1, 
\label{Gestimate1}
\\
 -1.46\times10^{-3} \le & \ a_s k_F \ & \le   -8.2, 
\label{asestimate1}
\eeq
where $r_m=m_1/m_2\simeq 4$ the mass ratio, $\lambda^2 \sim 1$. 
We have employed 
$k_F = \lk 6\pi^2 n_{\rm Dy}\rk^{1/3} \sim 1.81 \times 10^7 \, {\rm m^{-1}}$ 
and $-3000\, a_0\le a_{12}\le -40\, a_0$ with $a_0=5.29\times10^{-11}\, {\rm m}$ the Bohr radius, 
tunable from weak coupling to unitarity regime via interspecies Feshbach resonance \cite{Ravensbergen_2020}. 
As for the dipolar length $a_d=m_1d^2/3$, we find that $a_d\sim 131 a_0$ for the magnetic dipole moment of dysprosium 
$|\bd|\sim 10 \mu_B$ where $\mu_B$ is the Bohr magneton $\mu_B$ \cite{Chomaz_2022}. 

While these numerical values for $G$ and $a_s k_F$ 
are ballpark estimation as we have employed the Born approximation for $a_{12}$ and $a_s$, 
it seems quite possible that the effective interaction between impurity atoms are tunable in experiments 
from weak to strong coupling regime by means of adjusting parameters of surrounding dipolar Fermi gas. 
In this experimental setup, the deformation parameter is estimated to be $\beta = 1-\frac{2a_d k_F}{3\pi} \sim 0.97$, i.e., the deformation is just a little. 
To get smaller values of $\beta$, 
it is reasonable in experiment to increase $k_F$, i.e., 
the peak density, rather than $a_d$. 

\section{Summary and outlook} 
In summary, we have derived the induced interaction potential acting between two impurity atoms 
immersed in polarized dipolar Fermi gas at zero temperature, 
and found that it becomes an anisotropic RKKY potential. 
The anisotropy of the potential reflects the density fluctuation of the medium dipolar Fermi gas
under the influence of deformed Fermi surface.   
Then we have treated two-body problem of impurity atoms interacting with the induced potential. 
We have first solved the Schr\"{o}dinger equation to obtain eigenenergies of two impurity atoms, 
and determined a critical coupling strength above which the first bound state emerges. 
We have also investigated the scattering amplitude for two impurity atoms 
in the Born approximation 
to figure out its angle dependence with respect to dipole polarization direction,  
and made the partial-wave analysis to observe 
transitions between different angular momentum states in scattering processes. 

As outlook, it is interesting to extend the present two-body system of impurity atoms 
to a many-body one, 
since in reality impurity (minority) atoms usually exist as a dilute gas 
in the experiments for population imbalanced mixtures of majority and minority atoms. 
The extension involves effects of the particle statistics and self-energy of impurity atoms, 
and intraspecies interactions among impurity atoms such as the induced interaction obtained in this work, 
which is to be incorporated into, e.g., in-medium T matrix approach for the two-body correlation of impurity atoms 
\cite{Camacho_Guardian_2018,Moriya_2021,Tajima_2022,Muir_2022}. 

\begin{acknowledgments} 
This work was supported by Grants-in-Aid for Scientific
Research through Grant No.~21K03422, provided by JSPS. 
E.~N. is grateful to Shuichiro Ebata, Kei Iida, Hiroyuki Tajima, Kosai Tanabe, 
Atsushi Umeya, and Naotaka Yoshinaga 
for helpful discussion about in-medium few nucleon correlations. 
\end{acknowledgments} 

\bibliography{dipolebib.bib}

\begin{thebibliography}{66}%
\makeatletter
\providecommand \@ifxundefined [1]{%
 \@ifx{#1\undefined}
}%
\providecommand \@ifnum [1]{%
 \ifnum #1\expandafter \@firstoftwo
 \else \expandafter \@secondoftwo
 \fi
}%
\providecommand \@ifx [1]{%
 \ifx #1\expandafter \@firstoftwo
 \else \expandafter \@secondoftwo
 \fi
}%
\providecommand \natexlab [1]{#1}%
\providecommand \enquote  [1]{``#1''}%
\providecommand \bibnamefont  [1]{#1}%
\providecommand \bibfnamefont [1]{#1}%
\providecommand \citenamefont [1]{#1}%
\providecommand \href@noop [0]{\@secondoftwo}%
\providecommand \href [0]{\begingroup \@sanitize@url \@href}%
\providecommand \@href[1]{\@@startlink{#1}\@@href}%
\providecommand \@@href[1]{\endgroup#1\@@endlink}%
\providecommand \@sanitize@url [0]{\catcode `\\12\catcode `\$12\catcode
  `\&12\catcode `\#12\catcode `\^12\catcode `\_12\catcode `\%12\relax}%
\providecommand \@@startlink[1]{}%
\providecommand \@@endlink[0]{}%
\providecommand \url  [0]{\begingroup\@sanitize@url \@url }%
\providecommand \@url [1]{\endgroup\@href {#1}{\urlprefix }}%
\providecommand \urlprefix  [0]{URL }%
\providecommand \Eprint [0]{\href }%
\providecommand \doibase [0]{https://doi.org/}%
\providecommand \selectlanguage [0]{\@gobble}%
\providecommand \bibinfo  [0]{\@secondoftwo}%
\providecommand \bibfield  [0]{\@secondoftwo}%
\providecommand \translation [1]{[#1]}%
\providecommand \BibitemOpen [0]{}%
\providecommand \bibitemStop [0]{}%
\providecommand \bibitemNoStop [0]{.\EOS\space}%
\providecommand \EOS [0]{\spacefactor3000\relax}%
\providecommand \BibitemShut  [1]{\csname bibitem#1\endcsname}%
\let\auto@bib@innerbib\@empty
\bibitem [{\citenamefont {Kadau}\ \emph {et~al.}(2016)\citenamefont {Kadau},
  \citenamefont {Schmitt}, \citenamefont {Wenzel}, \citenamefont {Wink},
  \citenamefont {Maier}, \citenamefont {Ferrier-Barbut},\ and\ \citenamefont
  {Pfau}}]{Kadau_2016}%
  \BibitemOpen
  \bibfield  {author} {\bibinfo {author} {\bibfnamefont {H.}~\bibnamefont
  {Kadau}}, \bibinfo {author} {\bibfnamefont {M.}~\bibnamefont {Schmitt}},
  \bibinfo {author} {\bibfnamefont {M.}~\bibnamefont {Wenzel}}, \bibinfo
  {author} {\bibfnamefont {C.}~\bibnamefont {Wink}}, \bibinfo {author}
  {\bibfnamefont {T.}~\bibnamefont {Maier}}, \bibinfo {author} {\bibfnamefont
  {I.}~\bibnamefont {Ferrier-Barbut}},\ and\ \bibinfo {author} {\bibfnamefont
  {T.}~\bibnamefont {Pfau}},\ }\bibfield  {title} {\bibinfo {title} {Observing
  the rosensweig instability of a quantum ferrofluid},\ }\href
  {https://doi.org/10.1038/nature16485} {\bibfield  {journal} {\bibinfo
  {journal} {Nature}\ }\textbf {\bibinfo {volume} {530}},\ \bibinfo {pages}
  {194} (\bibinfo {year} {2016})}\BibitemShut {NoStop}%
\bibitem [{\citenamefont {Tanzi}\ \emph {et~al.}(2019)\citenamefont {Tanzi},
  \citenamefont {Lucioni}, \citenamefont {Fam{\`{a}}}, \citenamefont {Catani},
  \citenamefont {Fioretti}, \citenamefont {Gabbanini}, \citenamefont {Bisset},
  \citenamefont {Santos},\ and\ \citenamefont {Modugno}}]{Tanzi_2019}%
  \BibitemOpen
  \bibfield  {author} {\bibinfo {author} {\bibfnamefont {L.}~\bibnamefont
  {Tanzi}}, \bibinfo {author} {\bibfnamefont {E.}~\bibnamefont {Lucioni}},
  \bibinfo {author} {\bibfnamefont {F.}~\bibnamefont {Fam{\`{a}}}}, \bibinfo
  {author} {\bibfnamefont {J.}~\bibnamefont {Catani}}, \bibinfo {author}
  {\bibfnamefont {A.}~\bibnamefont {Fioretti}}, \bibinfo {author}
  {\bibfnamefont {C.}~\bibnamefont {Gabbanini}}, \bibinfo {author}
  {\bibfnamefont {R.~N.}\ \bibnamefont {Bisset}}, \bibinfo {author}
  {\bibfnamefont {L.}~\bibnamefont {Santos}},\ and\ \bibinfo {author}
  {\bibfnamefont {G.}~\bibnamefont {Modugno}},\ }\bibfield  {title} {\bibinfo
  {title} {Observation of a dipolar quantum gas with metastable supersolid
  properties},\ }\href {https://doi.org/10.1103/physrevlett.122.130405}
  {\bibfield  {journal} {\bibinfo  {journal} {Phys.~Rev.~Lett.}\ }\textbf
  {\bibinfo {volume} {122}},\ \bibinfo {pages} {130405} (\bibinfo {year}
  {2019})}\BibitemShut {NoStop}%
\bibitem [{\citenamefont {Hertkorn}\ \emph {et~al.}(2021)\citenamefont
  {Hertkorn}, \citenamefont {Schmidt}, \citenamefont {B\"{o}ttcher},
  \citenamefont {Guo}, \citenamefont {Schmidt}, \citenamefont {Ng},
  \citenamefont {Graham}, \citenamefont {B\"{o}chler}, \citenamefont {Langen},
  \citenamefont {Zwierlein},\ and\ \citenamefont {Pfau}}]{Hertkorn_2021}%
  \BibitemOpen
  \bibfield  {author} {\bibinfo {author} {\bibfnamefont {J.}~\bibnamefont
  {Hertkorn}}, \bibinfo {author} {\bibfnamefont {J.-N.}\ \bibnamefont
  {Schmidt}}, \bibinfo {author} {\bibfnamefont {F.}~\bibnamefont
  {B\"{o}ttcher}}, \bibinfo {author} {\bibfnamefont {M.}~\bibnamefont {Guo}},
  \bibinfo {author} {\bibfnamefont {M.}~\bibnamefont {Schmidt}}, \bibinfo
  {author} {\bibfnamefont {K.~S.~H.}\ \bibnamefont {Ng}}, \bibinfo {author}
  {\bibfnamefont {S.~D.}\ \bibnamefont {Graham}}, \bibinfo {author}
  {\bibfnamefont {H.~P.}\ \bibnamefont {B\"{o}chler}}, \bibinfo {author}
  {\bibfnamefont {T.}~\bibnamefont {Langen}}, \bibinfo {author} {\bibfnamefont
  {M.}~\bibnamefont {Zwierlein}},\ and\ \bibinfo {author} {\bibfnamefont
  {T.}~\bibnamefont {Pfau}},\ }\bibfield  {title} {\bibinfo {title} {Density
  fluctuations across the superfluid-supersolid phase transition in a dipolar
  quantum gas},\ }\href {https://doi.org/10.1103/physrevx.11.011037} {\bibfield
   {journal} {\bibinfo  {journal} {Physical Review X}\ }\textbf {\bibinfo
  {volume} {11}},\ \bibinfo {pages} {011037} (\bibinfo {year}
  {2021})}\BibitemShut {NoStop}%
\bibitem [{\citenamefont {Ferrier-Barbut}\ \emph {et~al.}(2016)\citenamefont
  {Ferrier-Barbut}, \citenamefont {Kadau}, \citenamefont {Schmitt},
  \citenamefont {Wenzel},\ and\ \citenamefont {Pfau}}]{Ferrier_Barbut_2016}%
  \BibitemOpen
  \bibfield  {author} {\bibinfo {author} {\bibfnamefont {I.}~\bibnamefont
  {Ferrier-Barbut}}, \bibinfo {author} {\bibfnamefont {H.}~\bibnamefont
  {Kadau}}, \bibinfo {author} {\bibfnamefont {M.}~\bibnamefont {Schmitt}},
  \bibinfo {author} {\bibfnamefont {M.}~\bibnamefont {Wenzel}},\ and\ \bibinfo
  {author} {\bibfnamefont {T.}~\bibnamefont {Pfau}},\ }\bibfield  {title}
  {\bibinfo {title} {Observation of quantum droplets in a strongly dipolar bose
  gas},\ }\href {https://doi.org/10.1103/physrevlett.116.215301} {\bibfield
  {journal} {\bibinfo  {journal} {Phys.~Rev.~Lett.}\ }\textbf {\bibinfo
  {volume} {116}},\ \bibinfo {pages} {215301} (\bibinfo {year}
  {2016})}\BibitemShut {NoStop}%
\bibitem [{\citenamefont {Fregoso}\ and\ \citenamefont
  {Fradkin}(2009)}]{Fregoso_2009b}%
  \BibitemOpen
  \bibfield  {author} {\bibinfo {author} {\bibfnamefont {B.~M.}\ \bibnamefont
  {Fregoso}}\ and\ \bibinfo {author} {\bibfnamefont {E.}~\bibnamefont
  {Fradkin}},\ }\bibfield  {title} {\bibinfo {title} {Ferronematic ground state
  of the dilute dipolar fermi gas},\ }\href
  {https://doi.org/10.1103/physrevlett.103.205301} {\bibfield  {journal}
  {\bibinfo  {journal} {Phys.~Rev.~Lett.}\ }\textbf {\bibinfo {volume} {103}},\
  \bibinfo {pages} {205301} (\bibinfo {year} {2009})}\BibitemShut {NoStop}%
\bibitem [{\citenamefont {Fregoso}\ and\ \citenamefont
  {Fradkin}(2010)}]{Fregoso_2010}%
  \BibitemOpen
  \bibfield  {author} {\bibinfo {author} {\bibfnamefont {B.~M.}\ \bibnamefont
  {Fregoso}}\ and\ \bibinfo {author} {\bibfnamefont {E.}~\bibnamefont
  {Fradkin}},\ }\bibfield  {title} {\bibinfo {title} {Unconventional magnetism
  in imbalanced fermi systems with magnetic dipolar interactions},\ }\href
  {https://doi.org/10.1103/physrevb.81.214443} {\bibfield  {journal} {\bibinfo
  {journal} {Physical Review B}\ }\textbf {\bibinfo {volume} {81}},\ \bibinfo
  {pages} {214443} (\bibinfo {year} {2010})}\BibitemShut {NoStop}%
\bibitem [{\citenamefont {Sogo}\ \emph {et~al.}(2012)\citenamefont {Sogo},
  \citenamefont {Urban}, \citenamefont {Schuck},\ and\ \citenamefont
  {Miyakawa}}]{Sogo_2012}%
  \BibitemOpen
  \bibfield  {author} {\bibinfo {author} {\bibfnamefont {T.}~\bibnamefont
  {Sogo}}, \bibinfo {author} {\bibfnamefont {M.}~\bibnamefont {Urban}},
  \bibinfo {author} {\bibfnamefont {P.}~\bibnamefont {Schuck}},\ and\ \bibinfo
  {author} {\bibfnamefont {T.}~\bibnamefont {Miyakawa}},\ }\bibfield  {title}
  {\bibinfo {title} {Spontaneous generation of spin-orbit coupling in magnetic
  dipolar fermi gases},\ }\href {https://doi.org/10.1103/PhysRevA.85.031601}
  {\bibfield  {journal} {\bibinfo  {journal} {Phys.~Rev.~A}\ }\textbf {\bibinfo
  {volume} {85}},\ \bibinfo {pages} {031601(R)} (\bibinfo {year}
  {2012})}\BibitemShut {NoStop}%
\bibitem [{\citenamefont {Bhongale}\ \emph {et~al.}(2013)\citenamefont
  {Bhongale}, \citenamefont {Mathey}, \citenamefont {Tsai}, \citenamefont
  {Clark},\ and\ \citenamefont {Zhao}}]{Bhongale_2013}%
  \BibitemOpen
  \bibfield  {author} {\bibinfo {author} {\bibfnamefont {S.~G.}\ \bibnamefont
  {Bhongale}}, \bibinfo {author} {\bibfnamefont {L.}~\bibnamefont {Mathey}},
  \bibinfo {author} {\bibfnamefont {S.-W.}\ \bibnamefont {Tsai}}, \bibinfo
  {author} {\bibfnamefont {C.~W.}\ \bibnamefont {Clark}},\ and\ \bibinfo
  {author} {\bibfnamefont {E.}~\bibnamefont {Zhao}},\ }\bibfield  {title}
  {\bibinfo {title} {Unconventional spin-density waves in dipolar fermi
  gases},\ }\href {https://doi.org/10.1103/physreva.87.043604} {\bibfield
  {journal} {\bibinfo  {journal} {Phys.~Rev.~A}\ }\textbf {\bibinfo {volume}
  {87}},\ \bibinfo {pages} {043604} (\bibinfo {year} {2013})}\BibitemShut
  {NoStop}%
\bibitem [{\citenamefont {Baranov}\ \emph {et~al.}(2002)\citenamefont
  {Baranov}, \citenamefont {Mar'enko}, \citenamefont {Rychkov},\ and\
  \citenamefont {Shlyapnikov}}]{Baranov_2002}%
  \BibitemOpen
  \bibfield  {author} {\bibinfo {author} {\bibfnamefont {M.~A.}\ \bibnamefont
  {Baranov}}, \bibinfo {author} {\bibfnamefont {M.~S.}\ \bibnamefont
  {Mar'enko}}, \bibinfo {author} {\bibfnamefont {V.~S.}\ \bibnamefont
  {Rychkov}},\ and\ \bibinfo {author} {\bibfnamefont {G.~V.}\ \bibnamefont
  {Shlyapnikov}},\ }\bibfield  {title} {\bibinfo {title} {Superfluid pairing in
  a polarized dipolar fermi gas},\ }\href
  {https://doi.org/10.1103/physreva.66.013606} {\bibfield  {journal} {\bibinfo
  {journal} {Phys.~Rev.~A}\ }\textbf {\bibinfo {volume} {66}},\ \bibinfo
  {pages} {013606} (\bibinfo {year} {2002})}\BibitemShut {NoStop}%
\bibitem [{\citenamefont {Baranov}\ \emph {et~al.}(2004)\citenamefont
  {Baranov}, \citenamefont {Dobrek},\ and\ \citenamefont
  {Lewenstein}}]{Baranov_2004}%
  \BibitemOpen
  \bibfield  {author} {\bibinfo {author} {\bibfnamefont {M.~A.}\ \bibnamefont
  {Baranov}}, \bibinfo {author} {\bibfnamefont {{\L}.}~\bibnamefont {Dobrek}},\
  and\ \bibinfo {author} {\bibfnamefont {M.}~\bibnamefont {Lewenstein}},\
  }\bibfield  {title} {\bibinfo {title} {Superfluidity of trapped dipolar fermi
  gases},\ }\href {https://doi.org/10.1103/physrevlett.92.250403} {\bibfield
  {journal} {\bibinfo  {journal} {Phys.~Rev.~Lett.}\ }\textbf {\bibinfo
  {volume} {92}},\ \bibinfo {pages} {250403} (\bibinfo {year}
  {2004})}\BibitemShut {NoStop}%
\bibitem [{\citenamefont {Bruun}\ and\ \citenamefont
  {Taylor}(2008)}]{Bruun_2008}%
  \BibitemOpen
  \bibfield  {author} {\bibinfo {author} {\bibfnamefont {G.~M.}\ \bibnamefont
  {Bruun}}\ and\ \bibinfo {author} {\bibfnamefont {E.}~\bibnamefont {Taylor}},\
  }\bibfield  {title} {\bibinfo {title} {Quantum phases of a two-dimensional
  dipolar fermi gas},\ }\href {https://doi.org/10.1103/physrevlett.101.245301}
  {\bibfield  {journal} {\bibinfo  {journal} {Phys.~Rev.~Lett.}\ }\textbf
  {\bibinfo {volume} {101}},\ \bibinfo {pages} {245301} (\bibinfo {year}
  {2008})}\BibitemShut {NoStop}%
\bibitem [{\citenamefont {Cooper}\ and\ \citenamefont
  {Shlyapnikov}(2009)}]{Cooper_2009}%
  \BibitemOpen
  \bibfield  {author} {\bibinfo {author} {\bibfnamefont {N.~R.}\ \bibnamefont
  {Cooper}}\ and\ \bibinfo {author} {\bibfnamefont {G.~V.}\ \bibnamefont
  {Shlyapnikov}},\ }\bibfield  {title} {\bibinfo {title} {Stable topological
  superfluid phase of ultracold polar fermionic molecules},\ }\href
  {https://doi.org/10.1103/physrevlett.103.155302} {\bibfield  {journal}
  {\bibinfo  {journal} {Phys.~Rev.~Lett.}\ }\textbf {\bibinfo {volume} {103}},\
  \bibinfo {pages} {155302} (\bibinfo {year} {2009})}\BibitemShut {NoStop}%
\bibitem [{\citenamefont {Levinsen}\ \emph {et~al.}(2011)\citenamefont
  {Levinsen}, \citenamefont {Cooper},\ and\ \citenamefont
  {Shlyapnikov}}]{Levinsen_2011}%
  \BibitemOpen
  \bibfield  {author} {\bibinfo {author} {\bibfnamefont {J.}~\bibnamefont
  {Levinsen}}, \bibinfo {author} {\bibfnamefont {N.~R.}\ \bibnamefont
  {Cooper}},\ and\ \bibinfo {author} {\bibfnamefont {G.~V.}\ \bibnamefont
  {Shlyapnikov}},\ }\bibfield  {title} {\bibinfo {title} {Topological
  superfluid phase of fermionic polar molecules},\ }\href
  {https://doi.org/10.1103/physreva.84.013603} {\bibfield  {journal} {\bibinfo
  {journal} {Phys.~Rev.~A}\ }\textbf {\bibinfo {volume} {84}},\ \bibinfo
  {pages} {013603} (\bibinfo {year} {2011})}\BibitemShut {NoStop}%
\bibitem [{\citenamefont {Griesmaier}\ \emph {et~al.}(2005)\citenamefont
  {Griesmaier}, \citenamefont {Wener}, \citenamefont {Hensler}, \citenamefont
  {Stuhler},\ and\ \citenamefont {Pfau}}]{Griesmaier_2005}%
  \BibitemOpen
  \bibfield  {author} {\bibinfo {author} {\bibfnamefont {A.}~\bibnamefont
  {Griesmaier}}, \bibinfo {author} {\bibfnamefont {J.}~\bibnamefont {Wener}},
  \bibinfo {author} {\bibfnamefont {S.}~\bibnamefont {Hensler}}, \bibinfo
  {author} {\bibfnamefont {J.}~\bibnamefont {Stuhler}},\ and\ \bibinfo {author}
  {\bibfnamefont {T.}~\bibnamefont {Pfau}},\ }\bibfield  {title} {\bibinfo
  {title} {Bose-einstein condensation of chromium},\ }\href
  {https://doi.org/https://doi.org/10.1103/PhysRevLett.94.160401} {\bibfield
  {journal} {\bibinfo  {journal} {Phys.~Rev.~Lett.}\ }\textbf {\bibinfo
  {volume} {94}},\ \bibinfo {pages} {160401} (\bibinfo {year}
  {2005})}\BibitemShut {NoStop}%
\bibitem [{\citenamefont {Lu}\ \emph {et~al.}(2011)\citenamefont {Lu},
  \citenamefont {Burdick}, \citenamefont {Youn},\ and\ \citenamefont
  {Lev}}]{Lu_2011}%
  \BibitemOpen
  \bibfield  {author} {\bibinfo {author} {\bibfnamefont {M.}~\bibnamefont
  {Lu}}, \bibinfo {author} {\bibfnamefont {N.~Q.}\ \bibnamefont {Burdick}},
  \bibinfo {author} {\bibfnamefont {S.~H.}\ \bibnamefont {Youn}},\ and\
  \bibinfo {author} {\bibfnamefont {B.~L.}\ \bibnamefont {Lev}},\ }\bibfield
  {title} {\bibinfo {title} {Strongly dipolar bose-einstein condensate of
  dysprosium},\ }\href
  {https://doi.org/https://doi.org/10.1103/PhysRevLett.107.190401} {\bibfield
  {journal} {\bibinfo  {journal} {Phys.~Rev.~Lett.}\ }\textbf {\bibinfo
  {volume} {107}},\ \bibinfo {pages} {190401} (\bibinfo {year}
  {2011})}\BibitemShut {NoStop}%
\bibitem [{\citenamefont {Aikawa}\ \emph {et~al.}(2012)\citenamefont {Aikawa},
  \citenamefont {Frisch}, \citenamefont {Mark}, \citenamefont {Baier},
  \citenamefont {Rietzler}, \citenamefont {Grimm}, ,\ and\ \citenamefont
  {Ferlaino}}]{Aikawa_2012}%
  \BibitemOpen
  \bibfield  {author} {\bibinfo {author} {\bibfnamefont {K.}~\bibnamefont
  {Aikawa}}, \bibinfo {author} {\bibfnamefont {A.}~\bibnamefont {Frisch}},
  \bibinfo {author} {\bibfnamefont {M.}~\bibnamefont {Mark}}, \bibinfo {author}
  {\bibfnamefont {S.}~\bibnamefont {Baier}}, \bibinfo {author} {\bibfnamefont
  {A.}~\bibnamefont {Rietzler}}, \bibinfo {author} {\bibfnamefont
  {R.}~\bibnamefont {Grimm}}, ,\ and\ \bibinfo {author} {\bibfnamefont
  {F.}~\bibnamefont {Ferlaino}},\ }\bibfield  {title} {\bibinfo {title}
  {Bose-einstein condensation of erbium},\ }\href
  {https://doi.org/https://doi.org/10.1103/PhysRevLett.108.210401} {\bibfield
  {journal} {\bibinfo  {journal} {Phys.~Rev.~Lett.}\ }\textbf {\bibinfo
  {volume} {108}},\ \bibinfo {pages} {210401} (\bibinfo {year}
  {2012})}\BibitemShut {NoStop}%
\bibitem [{\citenamefont {Lu}\ \emph {et~al.}(2012)\citenamefont {Lu},
  \citenamefont {Burdick},\ and\ \citenamefont {Lev}}]{Lu_2012}%
  \BibitemOpen
  \bibfield  {author} {\bibinfo {author} {\bibfnamefont {M.}~\bibnamefont
  {Lu}}, \bibinfo {author} {\bibfnamefont {N.~Q.}\ \bibnamefont {Burdick}},\
  and\ \bibinfo {author} {\bibfnamefont {B.~L.}\ \bibnamefont {Lev}},\
  }\bibfield  {title} {\bibinfo {title} {Quantum degenerate dipolar fermi
  gas},\ }\href {https://doi.org/10.1103/physrevlett.108.215301} {\bibfield
  {journal} {\bibinfo  {journal} {Phys.~Rev.~Lett.}\ }\textbf {\bibinfo
  {volume} {108}},\ \bibinfo {pages} {215301} (\bibinfo {year}
  {2012})}\BibitemShut {NoStop}%
\bibitem [{\citenamefont {Aikawa}\ \emph
  {et~al.}(2014{\natexlab{a}})\citenamefont {Aikawa}, \citenamefont {Frisch},
  \citenamefont {Mark}, \citenamefont {Baier}, \citenamefont {Grimm},\ and\
  \citenamefont {Ferlaino}}]{Aikawa_2014a}%
  \BibitemOpen
  \bibfield  {author} {\bibinfo {author} {\bibfnamefont {K.}~\bibnamefont
  {Aikawa}}, \bibinfo {author} {\bibfnamefont {A.}~\bibnamefont {Frisch}},
  \bibinfo {author} {\bibfnamefont {M.}~\bibnamefont {Mark}}, \bibinfo {author}
  {\bibfnamefont {S.}~\bibnamefont {Baier}}, \bibinfo {author} {\bibfnamefont
  {R.}~\bibnamefont {Grimm}},\ and\ \bibinfo {author} {\bibfnamefont
  {F.}~\bibnamefont {Ferlaino}},\ }\bibfield  {title} {\bibinfo {title}
  {Reaching fermi degeneracy via universal dipolar scattering},\ }\href
  {https://doi.org/https://doi.org/10.1103/PhysRevLett.112.010404} {\bibfield
  {journal} {\bibinfo  {journal} {Phys.~Rev.~Lett.}\ }\textbf {\bibinfo
  {volume} {112}},\ \bibinfo {pages} {010404} (\bibinfo {year}
  {2014}{\natexlab{a}})}\BibitemShut {NoStop}%
\bibitem [{\citenamefont {Naylor}\ \emph {et~al.}(2015)\citenamefont {Naylor},
  \citenamefont {Reigue}, \citenamefont {Mar{\'{e}}chal}, \citenamefont
  {Gorceix}, \citenamefont {Laburthe-Tolra},\ and\ \citenamefont
  {Vernac}}]{Naylor_2015}%
  \BibitemOpen
  \bibfield  {author} {\bibinfo {author} {\bibfnamefont {B.}~\bibnamefont
  {Naylor}}, \bibinfo {author} {\bibfnamefont {A.}~\bibnamefont {Reigue}},
  \bibinfo {author} {\bibfnamefont {E.}~\bibnamefont {Mar{\'{e}}chal}},
  \bibinfo {author} {\bibfnamefont {O.}~\bibnamefont {Gorceix}}, \bibinfo
  {author} {\bibfnamefont {B.}~\bibnamefont {Laburthe-Tolra}},\ and\ \bibinfo
  {author} {\bibfnamefont {L.}~\bibnamefont {Vernac}},\ }\bibfield  {title}
  {\bibinfo {title} {Chromium dipolar fermi sea},\ }\href
  {https://doi.org/https://doi.org/10.1103/PhysRevA.91.011603} {\bibfield
  {journal} {\bibinfo  {journal} {Phys.~Rev.~A.}\ }\textbf {\bibinfo {volume}
  {91}},\ \bibinfo {pages} {011603(R)} (\bibinfo {year} {2015})}\BibitemShut
  {NoStop}%
\bibitem [{\citenamefont {Baier}\ \emph {et~al.}(2018)\citenamefont {Baier},
  \citenamefont {Petter}, \citenamefont {Becher}, \citenamefont {Patscheider},
  \citenamefont {Natale}, \citenamefont {Chomaz}, \citenamefont {Mark},\ and\
  \citenamefont {Ferlaino}}]{Baier_2018}%
  \BibitemOpen
  \bibfield  {author} {\bibinfo {author} {\bibfnamefont {S.}~\bibnamefont
  {Baier}}, \bibinfo {author} {\bibfnamefont {D.}~\bibnamefont {Petter}},
  \bibinfo {author} {\bibfnamefont {J.~H.}\ \bibnamefont {Becher}}, \bibinfo
  {author} {\bibfnamefont {A.}~\bibnamefont {Patscheider}}, \bibinfo {author}
  {\bibfnamefont {G.}~\bibnamefont {Natale}}, \bibinfo {author} {\bibfnamefont
  {L.}~\bibnamefont {Chomaz}}, \bibinfo {author} {\bibfnamefont {M.~J.}\
  \bibnamefont {Mark}},\ and\ \bibinfo {author} {\bibfnamefont
  {F.}~\bibnamefont {Ferlaino}},\ }\bibfield  {title} {\bibinfo {title}
  {Realization of a strongly interacting fermi gas of dipolar atoms},\ }\href
  {https://doi.org/10.1103/physrevlett.121.093602} {\bibfield  {journal}
  {\bibinfo  {journal} {Phys.~Rev.~Lett.}\ }\textbf {\bibinfo {volume} {121}},\
  \bibinfo {pages} {093602} (\bibinfo {year} {2018})}\BibitemShut {NoStop}%
\bibitem [{\citenamefont {Durastante}\ \emph {et~al.}(2020)\citenamefont
  {Durastante}, \citenamefont {Politi}, \citenamefont {Sohmen}, \citenamefont
  {Ilzh\"{o}fer}, \citenamefont {Mark}, \citenamefont {Norcia},\ and\
  \citenamefont {Ferlaino}}]{Durastante_2020}%
  \BibitemOpen
  \bibfield  {author} {\bibinfo {author} {\bibfnamefont {G.}~\bibnamefont
  {Durastante}}, \bibinfo {author} {\bibfnamefont {C.}~\bibnamefont {Politi}},
  \bibinfo {author} {\bibfnamefont {M.}~\bibnamefont {Sohmen}}, \bibinfo
  {author} {\bibfnamefont {P.}~\bibnamefont {Ilzh\"{o}fer}}, \bibinfo {author}
  {\bibfnamefont {M.~J.}\ \bibnamefont {Mark}}, \bibinfo {author}
  {\bibfnamefont {M.~A.}\ \bibnamefont {Norcia}},\ and\ \bibinfo {author}
  {\bibfnamefont {F.}~\bibnamefont {Ferlaino}},\ }\bibfield  {title} {\bibinfo
  {title} {Feshbach resonances in an erbium-dysprosium dipolar mixture},\
  }\href {https://doi.org/10.1103/physreva.102.033330} {\bibfield  {journal}
  {\bibinfo  {journal} {Phys.~Rev.~A}\ }\textbf {\bibinfo {volume} {102}},\
  \bibinfo {pages} {033330} (\bibinfo {year} {2020})}\BibitemShut {NoStop}%
\bibitem [{\citenamefont {Chomaz}\ \emph {et~al.}(2022)\citenamefont {Chomaz},
  \citenamefont {Ferrier-Barbut}, \citenamefont {Ferlaino}, \citenamefont
  {Laburthe-Tolra}, \citenamefont {Lev},\ and\ \citenamefont
  {Pfau}}]{Chomaz_2022}%
  \BibitemOpen
  \bibfield  {author} {\bibinfo {author} {\bibfnamefont {L.}~\bibnamefont
  {Chomaz}}, \bibinfo {author} {\bibfnamefont {I.}~\bibnamefont
  {Ferrier-Barbut}}, \bibinfo {author} {\bibfnamefont {F.}~\bibnamefont
  {Ferlaino}}, \bibinfo {author} {\bibfnamefont {B.}~\bibnamefont
  {Laburthe-Tolra}}, \bibinfo {author} {\bibfnamefont {B.~L.}\ \bibnamefont
  {Lev}},\ and\ \bibinfo {author} {\bibfnamefont {T.}~\bibnamefont {Pfau}},\
  }\bibfield  {title} {\bibinfo {title} {Dipolar physics: a review of
  experiments with magnetic quantum gases},\ }\href
  {https://doi.org/10.1088/1361-6633/aca814} {\bibfield  {journal} {\bibinfo
  {journal} {Reports on Progress in Physics}\ }\textbf {\bibinfo {volume}
  {86}},\ \bibinfo {pages} {026401} (\bibinfo {year} {2022})}\BibitemShut
  {NoStop}%
\bibitem [{\citenamefont {Klaus}\ \emph {et~al.}(2022)\citenamefont {Klaus},
  \citenamefont {Bland}, \citenamefont {Poli}, \citenamefont {Politi},
  \citenamefont {Lamporesi}, \citenamefont {Casotti}, \citenamefont {Bisset},
  \citenamefont {Mark},\ and\ \citenamefont {Ferlaino}}]{Klaus_2022}%
  \BibitemOpen
  \bibfield  {author} {\bibinfo {author} {\bibfnamefont {L.}~\bibnamefont
  {Klaus}}, \bibinfo {author} {\bibfnamefont {T.}~\bibnamefont {Bland}},
  \bibinfo {author} {\bibfnamefont {E.}~\bibnamefont {Poli}}, \bibinfo {author}
  {\bibfnamefont {C.}~\bibnamefont {Politi}}, \bibinfo {author} {\bibfnamefont
  {G.}~\bibnamefont {Lamporesi}}, \bibinfo {author} {\bibfnamefont
  {E.}~\bibnamefont {Casotti}}, \bibinfo {author} {\bibfnamefont {R.~N.}\
  \bibnamefont {Bisset}}, \bibinfo {author} {\bibfnamefont {M.~J.}\
  \bibnamefont {Mark}},\ and\ \bibinfo {author} {\bibfnamefont
  {F.}~\bibnamefont {Ferlaino}},\ }\bibfield  {title} {\bibinfo {title}
  {Observation of vortices and vortex stripes in a dipolar condensate},\ }\href
  {https://doi.org/10.1038/s41567-022-01793-8} {\bibfield  {journal} {\bibinfo
  {journal} {Nature Physics}\ }\textbf {\bibinfo {volume} {18}},\ \bibinfo
  {pages} {1453} (\bibinfo {year} {2022})}\BibitemShut {NoStop}%
\bibitem [{\citenamefont {Scheiermann}\ \emph {et~al.}(2023)\citenamefont
  {Scheiermann}, \citenamefont {Ardila}, \citenamefont {Bland}, \citenamefont
  {Bisset},\ and\ \citenamefont {Santos}}]{Scheiermann_2023}%
  \BibitemOpen
  \bibfield  {author} {\bibinfo {author} {\bibfnamefont {D.}~\bibnamefont
  {Scheiermann}}, \bibinfo {author} {\bibfnamefont {L.~A.~P.}\ \bibnamefont
  {Ardila}}, \bibinfo {author} {\bibfnamefont {T.}~\bibnamefont {Bland}},
  \bibinfo {author} {\bibfnamefont {R.~N.}\ \bibnamefont {Bisset}},\ and\
  \bibinfo {author} {\bibfnamefont {L.}~\bibnamefont {Santos}},\ }\bibfield
  {title} {\bibinfo {title} {Catalyzation of supersolidity in binary dipolar
  condensates},\ }\href {https://doi.org/10.1103/physreva.107.l021302}
  {\bibfield  {journal} {\bibinfo  {journal} {Phys.~Rev.~A}\ }\textbf {\bibinfo
  {volume} {107}},\ \bibinfo {pages} {l021302} (\bibinfo {year}
  {2023})}\BibitemShut {NoStop}%
\bibitem [{\citenamefont {Petrov}(2015)}]{Petrov_2015}%
  \BibitemOpen
  \bibfield  {author} {\bibinfo {author} {\bibfnamefont {D.~S.}\ \bibnamefont
  {Petrov}},\ }\bibfield  {title} {\bibinfo {title} {Quantum mechanical
  stabilization of a collapsing bose-bose mixture},\ }\href
  {https://doi.org/10.1103/physrevlett.115.155302} {\bibfield  {journal}
  {\bibinfo  {journal} {Phys.~Rev.~Lett.}\ }\textbf {\bibinfo {volume} {115}},\
  \bibinfo {pages} {155302} (\bibinfo {year} {2015})}\BibitemShut {NoStop}%
\bibitem [{\citenamefont {Kumar}\ \emph {et~al.}(2017)\citenamefont {Kumar},
  \citenamefont {Muruganandam}, \citenamefont {Tomio},\ and\ \citenamefont
  {Gammal}}]{Kumar_2017}%
  \BibitemOpen
  \bibfield  {author} {\bibinfo {author} {\bibfnamefont {R.~K.}\ \bibnamefont
  {Kumar}}, \bibinfo {author} {\bibfnamefont {P.}~\bibnamefont {Muruganandam}},
  \bibinfo {author} {\bibfnamefont {L.}~\bibnamefont {Tomio}},\ and\ \bibinfo
  {author} {\bibfnamefont {A.}~\bibnamefont {Gammal}},\ }\bibfield  {title}
  {\bibinfo {title} {Miscibility in coupled dipolar and non-dipolar
  bose-einstein condensates},\ }\href
  {https://doi.org/10.1088/2399-6528/aa8db5} {\bibfield  {journal} {\bibinfo
  {journal} {Journal of Physics Communications}\ }\textbf {\bibinfo {volume}
  {1}},\ \bibinfo {pages} {035012} (\bibinfo {year} {2017})}\BibitemShut
  {NoStop}%
\bibitem [{\citenamefont {Miyakawa}\ \emph {et~al.}(2008)\citenamefont
  {Miyakawa}, \citenamefont {Sogo},\ and\ \citenamefont {Pu}}]{Miyakawa_2008}%
  \BibitemOpen
  \bibfield  {author} {\bibinfo {author} {\bibfnamefont {T.}~\bibnamefont
  {Miyakawa}}, \bibinfo {author} {\bibfnamefont {T.}~\bibnamefont {Sogo}},\
  and\ \bibinfo {author} {\bibfnamefont {H.}~\bibnamefont {Pu}},\ }\bibfield
  {title} {\bibinfo {title} {Phase-space deformation of a trapped dipolar fermi
  gas},\ }\href {https://doi.org/10.1103/physreva.77.061603} {\bibfield
  {journal} {\bibinfo  {journal} {Phys.~Rev.~A}\ }\textbf {\bibinfo {volume}
  {77}},\ \bibinfo {pages} {061603} (\bibinfo {year} {2008})}\BibitemShut
  {NoStop}%
\bibitem [{\citenamefont {Sogo}\ \emph {et~al.}(2009)\citenamefont {Sogo},
  \citenamefont {He}, \citenamefont {Miyakawa}, \citenamefont {Yi},
  \citenamefont {Lu},\ and\ \citenamefont {Pu}}]{Sogo_2009}%
  \BibitemOpen
  \bibfield  {author} {\bibinfo {author} {\bibfnamefont {T.}~\bibnamefont
  {Sogo}}, \bibinfo {author} {\bibfnamefont {L.}~\bibnamefont {He}}, \bibinfo
  {author} {\bibfnamefont {T.}~\bibnamefont {Miyakawa}}, \bibinfo {author}
  {\bibfnamefont {S.}~\bibnamefont {Yi}}, \bibinfo {author} {\bibfnamefont
  {H.}~\bibnamefont {Lu}},\ and\ \bibinfo {author} {\bibfnamefont
  {H.}~\bibnamefont {Pu}},\ }\bibfield  {title} {\bibinfo {title} {Dynamical
  properties of dipolar fermi gases},\ }\href
  {https://doi.org/10.1088/1367-2630/11/5/055017} {\bibfield  {journal}
  {\bibinfo  {journal} {New Journal of Physics}\ }\textbf {\bibinfo {volume}
  {11}},\ \bibinfo {pages} {055017} (\bibinfo {year} {2009})}\BibitemShut
  {NoStop}%
\bibitem [{\citenamefont {Tohyama}(2009)}]{Tohyama_2009}%
  \BibitemOpen
  \bibfield  {author} {\bibinfo {author} {\bibfnamefont {M.}~\bibnamefont
  {Tohyama}},\ }\bibfield  {title} {\bibinfo {title} {Quantum study of a
  trapped dipolar fermi gas},\ }\href
  {https://doi.org/https://doi.org/10.1143/JPSJ.78.104003} {\bibfield
  {journal} {\bibinfo  {journal} {Journal of the Physical Society of Japan}\
  }\textbf {\bibinfo {volume} {78}},\ \bibinfo {pages} {104003} (\bibinfo
  {year} {2009})}\BibitemShut {NoStop}%
\bibitem [{\citenamefont {Zhang}\ and\ \citenamefont {Yi}(2010)}]{Zhang_2010}%
  \BibitemOpen
  \bibfield  {author} {\bibinfo {author} {\bibfnamefont {J.-N.}\ \bibnamefont
  {Zhang}}\ and\ \bibinfo {author} {\bibfnamefont {S.}~\bibnamefont {Yi}},\
  }\bibfield  {title} {\bibinfo {title} {Thermodynamic properties of a dipolar
  fermi gas},\ }\href {https://doi.org/10.1103/physreva.81.033617} {\bibfield
  {journal} {\bibinfo  {journal} {Phys.~Rev.~A}\ }\textbf {\bibinfo {volume}
  {81}},\ \bibinfo {pages} {033617} (\bibinfo {year} {2010})}\BibitemShut
  {NoStop}%
\bibitem [{\citenamefont {Parish}\ and\ \citenamefont
  {Marchetti}(2012)}]{Parish_2012}%
  \BibitemOpen
  \bibfield  {author} {\bibinfo {author} {\bibfnamefont {M.~M.}\ \bibnamefont
  {Parish}}\ and\ \bibinfo {author} {\bibfnamefont {F.~M.}\ \bibnamefont
  {Marchetti}},\ }\bibfield  {title} {\bibinfo {title} {Density instabilities
  in a two-dimensional dipolar fermi gas},\ }\href
  {https://doi.org/10.1103/physrevlett.108.145304} {\bibfield  {journal}
  {\bibinfo  {journal} {Phys.~Rev.~Lett.}\ }\textbf {\bibinfo {volume} {108}},\
  \bibinfo {pages} {145304} (\bibinfo {year} {2012})}\BibitemShut {NoStop}%
\bibitem [{\citenamefont {Zhou}\ \emph {et~al.}(2021)\citenamefont {Zhou},
  \citenamefont {Nath}, \citenamefont {Wu}, \citenamefont {Lesanovsky},\ and\
  \citenamefont {Li}}]{Zhou_2021}%
  \BibitemOpen
  \bibfield  {author} {\bibinfo {author} {\bibfnamefont {Y.}~\bibnamefont
  {Zhou}}, \bibinfo {author} {\bibfnamefont {R.}~\bibnamefont {Nath}}, \bibinfo
  {author} {\bibfnamefont {H.}~\bibnamefont {Wu}}, \bibinfo {author}
  {\bibfnamefont {I.}~\bibnamefont {Lesanovsky}},\ and\ \bibinfo {author}
  {\bibfnamefont {W.}~\bibnamefont {Li}},\ }\bibfield  {title} {\bibinfo
  {title} {Multipolar fermi-surface deformation in a rydberg-dressed fermi gas
  with long-range anisotropic interactions},\ }\href
  {https://doi.org/10.1103/physreva.104.l061302} {\bibfield  {journal}
  {\bibinfo  {journal} {Phys.~Rev.~A}\ }\textbf {\bibinfo {volume} {104}},\
  \bibinfo {pages} {l061302} (\bibinfo {year} {2021})}\BibitemShut {NoStop}%
\bibitem [{\citenamefont {Chan}\ \emph {et~al.}(2010)\citenamefont {Chan},
  \citenamefont {Wu}, \citenamefont {Lee},\ and\ \citenamefont
  {Sarma}}]{Chan_2010}%
  \BibitemOpen
  \bibfield  {author} {\bibinfo {author} {\bibfnamefont {C.-K.}\ \bibnamefont
  {Chan}}, \bibinfo {author} {\bibfnamefont {C.}~\bibnamefont {Wu}}, \bibinfo
  {author} {\bibfnamefont {W.-C.}\ \bibnamefont {Lee}},\ and\ \bibinfo {author}
  {\bibfnamefont {S.~D.}\ \bibnamefont {Sarma}},\ }\bibfield  {title} {\bibinfo
  {title} {Anisotropic-fermi-liquid theory of ultracold fermionic polar
  molecules: Landau parameters and collective modes},\ }\href
  {https://doi.org/10.1103/physreva.81.023602} {\bibfield  {journal} {\bibinfo
  {journal} {Phys.~Rev.~A}\ }\textbf {\bibinfo {volume} {81}},\ \bibinfo
  {pages} {023602} (\bibinfo {year} {2010})}\BibitemShut {NoStop}%
\bibitem [{\citenamefont {Ronen}\ and\ \citenamefont
  {Bohn}(2010)}]{Ronen_2010}%
  \BibitemOpen
  \bibfield  {author} {\bibinfo {author} {\bibfnamefont {S.}~\bibnamefont
  {Ronen}}\ and\ \bibinfo {author} {\bibfnamefont {J.~L.}\ \bibnamefont
  {Bohn}},\ }\bibfield  {title} {\bibinfo {title} {Zero sound in dipolar fermi
  gases},\ }\href {https://doi.org/10.1103/physreva.81.033601} {\bibfield
  {journal} {\bibinfo  {journal} {Phys.~Rev.~A}\ }\textbf {\bibinfo {volume}
  {81}},\ \bibinfo {pages} {033601} (\bibinfo {year} {2010})}\BibitemShut
  {NoStop}%
\bibitem [{\citenamefont {Sieberer}\ and\ \citenamefont
  {Baranov}(2011)}]{Sieberer_2011}%
  \BibitemOpen
  \bibfield  {author} {\bibinfo {author} {\bibfnamefont {L.~M.}\ \bibnamefont
  {Sieberer}}\ and\ \bibinfo {author} {\bibfnamefont {M.~A.}\ \bibnamefont
  {Baranov}},\ }\bibfield  {title} {\bibinfo {title} {Collective modes,
  stability, and superfluid transition of a quasi-two-dimensional dipolar fermi
  gas},\ }\href {https://doi.org/10.1103/physreva.84.063633} {\bibfield
  {journal} {\bibinfo  {journal} {Phys.~Rev.~A}\ }\textbf {\bibinfo {volume}
  {84}},\ \bibinfo {pages} {063633} (\bibinfo {year} {2011})}\BibitemShut
  {NoStop}%
\bibitem [{\citenamefont {Miyakawa}\ \emph {et~al.}(2023)\citenamefont
  {Miyakawa}, \citenamefont {Nakano},\ and\ \citenamefont {Yabu}}]{MNY_2023}%
  \BibitemOpen
  \bibfield  {author} {\bibinfo {author} {\bibfnamefont {T.}~\bibnamefont
  {Miyakawa}}, \bibinfo {author} {\bibfnamefont {E.}~\bibnamefont {Nakano}},\
  and\ \bibinfo {author} {\bibfnamefont {H.}~\bibnamefont {Yabu}},\ }\bibfield
  {title} {\bibinfo {title} {Collective excitation modes in a dipolar and
  non-dipolar fermi gas mixture},\ }\href
  {https://doi.org/https://doi.org/10.7566/JPSCP.38.011014} {\bibfield
  {journal} {\bibinfo  {journal} {JPS Conf. Proc.}\ }\textbf {\bibinfo {volume}
  {38}},\ \bibinfo {pages} {011014} (\bibinfo {year} {2023})}\BibitemShut
  {NoStop}%
\bibitem [{\citenamefont {Aikawa}\ \emph
  {et~al.}(2014{\natexlab{b}})\citenamefont {Aikawa}, \citenamefont {Baier},
  \citenamefont {Frisch}, \citenamefont {Mark}, \citenamefont {Ravensbergen},\
  and\ \citenamefont {Ferlaino}}]{Aikawa_2014b}%
  \BibitemOpen
  \bibfield  {author} {\bibinfo {author} {\bibfnamefont {K.}~\bibnamefont
  {Aikawa}}, \bibinfo {author} {\bibfnamefont {S.}~\bibnamefont {Baier}},
  \bibinfo {author} {\bibfnamefont {A.}~\bibnamefont {Frisch}}, \bibinfo
  {author} {\bibfnamefont {M.}~\bibnamefont {Mark}}, \bibinfo {author}
  {\bibfnamefont {C.}~\bibnamefont {Ravensbergen}},\ and\ \bibinfo {author}
  {\bibfnamefont {F.}~\bibnamefont {Ferlaino}},\ }\bibfield  {title} {\bibinfo
  {title} {Observation of fermi surface deformation in a dipolar quantum gas},\
  }\href {https://doi.org/10.1126/science.1255259} {\bibfield  {journal}
  {\bibinfo  {journal} {Science}\ }\textbf {\bibinfo {volume} {345}},\ \bibinfo
  {pages} {1484} (\bibinfo {year} {2014}{\natexlab{b}})}\BibitemShut {NoStop}%
\bibitem [{\citenamefont {Fregoso}\ \emph {et~al.}(2009)\citenamefont
  {Fregoso}, \citenamefont {Sun}, \citenamefont {Fradkin},\ and\ \citenamefont
  {Lev}}]{Fregoso_2009}%
  \BibitemOpen
  \bibfield  {author} {\bibinfo {author} {\bibfnamefont {B.~M.}\ \bibnamefont
  {Fregoso}}, \bibinfo {author} {\bibfnamefont {K.}~\bibnamefont {Sun}},
  \bibinfo {author} {\bibfnamefont {E.}~\bibnamefont {Fradkin}},\ and\ \bibinfo
  {author} {\bibfnamefont {B.~L.}\ \bibnamefont {Lev}},\ }\bibfield  {title}
  {\bibinfo {title} {Biaxial nematic phases in ultracold dipolar fermi gases},\
  }\href {https://doi.org/10.1088/1367-2630/11/10/103003} {\bibfield  {journal}
  {\bibinfo  {journal} {New Journal of Physics}\ }\textbf {\bibinfo {volume}
  {11}},\ \bibinfo {pages} {103003} (\bibinfo {year} {2009})}\BibitemShut
  {NoStop}%
\bibitem [{\citenamefont {Aikawa}\ \emph
  {et~al.}(2014{\natexlab{c}})\citenamefont {Aikawa}, \citenamefont {Frisch},
  \citenamefont {Mark}, \citenamefont {Baier}, \citenamefont {Grimm},
  \citenamefont {Bohn}, \citenamefont {Jin}, \citenamefont {Bruun},\ and\
  \citenamefont {Ferlaino}}]{Aikawa_2014c}%
  \BibitemOpen
  \bibfield  {author} {\bibinfo {author} {\bibfnamefont {K.}~\bibnamefont
  {Aikawa}}, \bibinfo {author} {\bibfnamefont {A.}~\bibnamefont {Frisch}},
  \bibinfo {author} {\bibfnamefont {M.}~\bibnamefont {Mark}}, \bibinfo {author}
  {\bibfnamefont {S.}~\bibnamefont {Baier}}, \bibinfo {author} {\bibfnamefont
  {R.}~\bibnamefont {Grimm}}, \bibinfo {author} {\bibfnamefont {J.~L.}\
  \bibnamefont {Bohn}}, \bibinfo {author} {\bibfnamefont {D.~S.}\ \bibnamefont
  {Jin}}, \bibinfo {author} {\bibfnamefont {G.~M.}\ \bibnamefont {Bruun}},\
  and\ \bibinfo {author} {\bibfnamefont {F.}~\bibnamefont {Ferlaino}},\
  }\bibfield  {title} {\bibinfo {title} {Anisotropic relaxation dynamics in a
  dipolar fermi gas driven out of equilibrium},\ }\href
  {https://doi.org/10.1103/physrevlett.113.263201} {\bibfield  {journal}
  {\bibinfo  {journal} {Phys.~Rev.~Lett.}\ }\textbf {\bibinfo {volume} {113}},\
  \bibinfo {pages} {263201} (\bibinfo {year} {2014}{\natexlab{c}})}\BibitemShut
  {NoStop}%
\bibitem [{\citenamefont {Wang}\ and\ \citenamefont {Bohn}(2021)}]{Wang_2021}%
  \BibitemOpen
  \bibfield  {author} {\bibinfo {author} {\bibfnamefont {R.~R.~W.}\
  \bibnamefont {Wang}}\ and\ \bibinfo {author} {\bibfnamefont {J.~L.}\
  \bibnamefont {Bohn}},\ }\bibfield  {title} {\bibinfo {title} {Anisotropic
  thermalization of dilute dipolar gases},\ }\href
  {https://doi.org/10.1103/physreva.103.063320} {\bibfield  {journal} {\bibinfo
   {journal} {Phys.~Rev.~A}\ }\textbf {\bibinfo {volume} {103}},\ \bibinfo
  {pages} {063320} (\bibinfo {year} {2021})}\BibitemShut {NoStop}%
\bibitem [{\citenamefont {Ravensbergen}\ \emph {et~al.}(2018)\citenamefont
  {Ravensbergen}, \citenamefont {Corre}, \citenamefont {Soave}, \citenamefont
  {Kreyer}, \citenamefont {Kirilov},\ and\ \citenamefont
  {Grimm}}]{Ravensbergen_2018}%
  \BibitemOpen
  \bibfield  {author} {\bibinfo {author} {\bibfnamefont {C.}~\bibnamefont
  {Ravensbergen}}, \bibinfo {author} {\bibfnamefont {V.}~\bibnamefont {Corre}},
  \bibinfo {author} {\bibfnamefont {E.}~\bibnamefont {Soave}}, \bibinfo
  {author} {\bibfnamefont {M.}~\bibnamefont {Kreyer}}, \bibinfo {author}
  {\bibfnamefont {E.}~\bibnamefont {Kirilov}},\ and\ \bibinfo {author}
  {\bibfnamefont {R.}~\bibnamefont {Grimm}},\ }\bibfield  {title} {\bibinfo
  {title} {Production of a degenerate fermi-fermi mixture of dysprosium and
  potassium atoms},\ }\href {https://doi.org/10.1103/physreva.98.063624}
  {\bibfield  {journal} {\bibinfo  {journal} {Phys.~Rev.~A}\ }\textbf {\bibinfo
  {volume} {98}},\ \bibinfo {pages} {063624} (\bibinfo {year}
  {2018})}\BibitemShut {NoStop}%
\bibitem [{\citenamefont {Ravensbergen}\ \emph {et~al.}(2020)\citenamefont
  {Ravensbergen}, \citenamefont {Soave}, \citenamefont {Corre}, \citenamefont
  {Kreyer}, \citenamefont {Huang}, \citenamefont {Kirilov},\ and\ \citenamefont
  {Grimm}}]{Ravensbergen_2020}%
  \BibitemOpen
  \bibfield  {author} {\bibinfo {author} {\bibfnamefont {C.}~\bibnamefont
  {Ravensbergen}}, \bibinfo {author} {\bibfnamefont {E.}~\bibnamefont {Soave}},
  \bibinfo {author} {\bibfnamefont {V.}~\bibnamefont {Corre}}, \bibinfo
  {author} {\bibfnamefont {M.}~\bibnamefont {Kreyer}}, \bibinfo {author}
  {\bibfnamefont {B.}~\bibnamefont {Huang}}, \bibinfo {author} {\bibfnamefont
  {E.}~\bibnamefont {Kirilov}},\ and\ \bibinfo {author} {\bibfnamefont
  {R.}~\bibnamefont {Grimm}},\ }\bibfield  {title} {\bibinfo {title}
  {Resonantly interacting fermi-fermi mixture of and},\ }\href
  {https://doi.org/10.1103/physrevlett.124.203402} {\bibfield  {journal}
  {\bibinfo  {journal} {Phys.~Rev.~Lett.}\ }\textbf {\bibinfo {volume} {124}},\
  \bibinfo {pages} {203402} (\bibinfo {year} {2020})}\BibitemShut {NoStop}%
\bibitem [{\citenamefont {Ye}\ \emph {et~al.}(2022)\citenamefont {Ye},
  \citenamefont {Canali}, \citenamefont {Soave}, \citenamefont {Kreyer},
  \citenamefont {Yudkin}, \citenamefont {Ravensbergen}, \citenamefont
  {Kirilov},\ and\ \citenamefont {Grimm}}]{Ye_2022}%
  \BibitemOpen
  \bibfield  {author} {\bibinfo {author} {\bibfnamefont {Z.-X.}\ \bibnamefont
  {Ye}}, \bibinfo {author} {\bibfnamefont {A.}~\bibnamefont {Canali}}, \bibinfo
  {author} {\bibfnamefont {E.}~\bibnamefont {Soave}}, \bibinfo {author}
  {\bibfnamefont {M.}~\bibnamefont {Kreyer}}, \bibinfo {author} {\bibfnamefont
  {Y.}~\bibnamefont {Yudkin}}, \bibinfo {author} {\bibfnamefont
  {C.}~\bibnamefont {Ravensbergen}}, \bibinfo {author} {\bibfnamefont
  {E.}~\bibnamefont {Kirilov}},\ and\ \bibinfo {author} {\bibfnamefont
  {R.}~\bibnamefont {Grimm}},\ }\bibfield  {title} {\bibinfo {title}
  {Observation of low-field feshbach resonances between and},\ }\href
  {https://doi.org/10.1103/physreva.106.043314} {\bibfield  {journal} {\bibinfo
   {journal} {Phys.~Rev.~A}\ }\textbf {\bibinfo {volume} {106}},\ \bibinfo
  {pages} {043314} (\bibinfo {year} {2022})}\BibitemShut {NoStop}%
\bibitem [{\citenamefont {Ciamei}\ \emph
  {et~al.}(2022{\natexlab{a}})\citenamefont {Ciamei}, \citenamefont {Finelli},
  \citenamefont {Trenkwalder}, \citenamefont {Inguscio}, \citenamefont
  {Simoni},\ and\ \citenamefont {Zaccanti}}]{Ciamei_2022}%
  \BibitemOpen
  \bibfield  {author} {\bibinfo {author} {\bibfnamefont {A.}~\bibnamefont
  {Ciamei}}, \bibinfo {author} {\bibfnamefont {S.}~\bibnamefont {Finelli}},
  \bibinfo {author} {\bibfnamefont {A.}~\bibnamefont {Trenkwalder}}, \bibinfo
  {author} {\bibfnamefont {M.}~\bibnamefont {Inguscio}}, \bibinfo {author}
  {\bibfnamefont {A.}~\bibnamefont {Simoni}},\ and\ \bibinfo {author}
  {\bibfnamefont {M.}~\bibnamefont {Zaccanti}},\ }\bibfield  {title} {\bibinfo
  {title} {Exploring ultracold collisions in fermi mixtures: Feshbach
  resonances and scattering properties of a novel alkali-transition metal
  system},\ }\href {https://doi.org/10.1103/physrevlett.129.093402} {\bibfield
  {journal} {\bibinfo  {journal} {Phys.~Rev.~Lett.}\ }\textbf {\bibinfo
  {volume} {129}},\ \bibinfo {pages} {093402} (\bibinfo {year}
  {2022}{\natexlab{a}})}\BibitemShut {NoStop}%
\bibitem [{\citenamefont {Ciamei}\ \emph
  {et~al.}(2022{\natexlab{b}})\citenamefont {Ciamei}, \citenamefont {Finelli},
  \citenamefont {Cosco}, \citenamefont {Inguscio}, \citenamefont
  {Trenkwalder},\ and\ \citenamefont {Zaccanti}}]{Ciamei_2022_2}%
  \BibitemOpen
  \bibfield  {author} {\bibinfo {author} {\bibfnamefont {A.}~\bibnamefont
  {Ciamei}}, \bibinfo {author} {\bibfnamefont {S.}~\bibnamefont {Finelli}},
  \bibinfo {author} {\bibfnamefont {A.}~\bibnamefont {Cosco}}, \bibinfo
  {author} {\bibfnamefont {M.}~\bibnamefont {Inguscio}}, \bibinfo {author}
  {\bibfnamefont {A.}~\bibnamefont {Trenkwalder}},\ and\ \bibinfo {author}
  {\bibfnamefont {M.}~\bibnamefont {Zaccanti}},\ }\bibfield  {title} {\bibinfo
  {title} {Double-degenerate fermi mixtures of 53cr and 6li atoms},\ }\href
  {https://doi.org/10.1103/physreva.106.053318} {\bibfield  {journal} {\bibinfo
   {journal} {Phys.~Rev.~A}\ }\textbf {\bibinfo {volume} {106}},\ \bibinfo
  {pages} {053318} (\bibinfo {year} {2022}{\natexlab{b}})}\BibitemShut
  {NoStop}%
\bibitem [{\citenamefont {Sch\"{a}fer}\ \emph {et~al.}(2022)\citenamefont
  {Sch\"{a}fer}, \citenamefont {Mizukami},\ and\ \citenamefont
  {Takahashi}}]{Sch_fer_2022}%
  \BibitemOpen
  \bibfield  {author} {\bibinfo {author} {\bibfnamefont {F.}~\bibnamefont
  {Sch\"{a}fer}}, \bibinfo {author} {\bibfnamefont {N.}~\bibnamefont
  {Mizukami}},\ and\ \bibinfo {author} {\bibfnamefont {Y.}~\bibnamefont
  {Takahashi}},\ }\bibfield  {title} {\bibinfo {title} {Feshbach resonances of
  large-mass-imbalance er-li mixtures},\ }\href
  {https://doi.org/10.1103/physreva.105.012816} {\bibfield  {journal} {\bibinfo
   {journal} {Phys.~Rev.~A}\ }\textbf {\bibinfo {volume} {105}},\ \bibinfo
  {pages} {012816} (\bibinfo {year} {2022})}\BibitemShut {NoStop}%
\bibitem [{\citenamefont {Sch\"{a}fer}\ \emph
  {et~al.}(2023{\natexlab{a}})\citenamefont {Sch\"{a}fer}, \citenamefont
  {Haruna},\ and\ \citenamefont {Takahashi}}]{Sch_fer_2023}%
  \BibitemOpen
  \bibfield  {author} {\bibinfo {author} {\bibfnamefont {F.}~\bibnamefont
  {Sch\"{a}fer}}, \bibinfo {author} {\bibfnamefont {Y.}~\bibnamefont
  {Haruna}},\ and\ \bibinfo {author} {\bibfnamefont {Y.}~\bibnamefont
  {Takahashi}},\ }\bibfield  {title} {\bibinfo {title} {Observation of feshbach
  resonances in an er-li fermi-fermi mixture},\ }\href
  {https://doi.org/10.7566/jpsj.92.054301} {\bibfield  {journal} {\bibinfo
  {journal} {Journal of the Physical Society of Japan}\ }\textbf {\bibinfo
  {volume} {92}},\ \bibinfo {pages} {054301} (\bibinfo {year}
  {2023}{\natexlab{a}})}\BibitemShut {NoStop}%
\bibitem [{\citenamefont {Sch\"{a}fer}\ \emph
  {et~al.}(2023{\natexlab{b}})\citenamefont {Sch\"{a}fer}, \citenamefont
  {Haruna},\ and\ \citenamefont {Takahashi}}]{Sch_fer_2023_2}%
  \BibitemOpen
  \bibfield  {author} {\bibinfo {author} {\bibfnamefont {F.}~\bibnamefont
  {Sch\"{a}fer}}, \bibinfo {author} {\bibfnamefont {Y.}~\bibnamefont
  {Haruna}},\ and\ \bibinfo {author} {\bibfnamefont {Y.}~\bibnamefont
  {Takahashi}},\ }\bibfield  {title} {\bibinfo {title} {Realization of a
  quantum degenerate mixture of highly magnetic and nonmagnetic atoms},\ }\href
  {https://doi.org/10.1103/physreva.107.l031306} {\bibfield  {journal}
  {\bibinfo  {journal} {Phys.~Rev.~A}\ }\textbf {\bibinfo {volume} {107}},\
  \bibinfo {pages} {l031306} (\bibinfo {year}
  {2023}{\natexlab{b}})}\BibitemShut {NoStop}%
\bibitem [{\citenamefont {Baarsma}\ and\ \citenamefont
  {T\"{o}rm\"{a}}(2017)}]{baarsma2017fermi}%
  \BibitemOpen
  \bibfield  {author} {\bibinfo {author} {\bibfnamefont {J.~E.}\ \bibnamefont
  {Baarsma}}\ and\ \bibinfo {author} {\bibfnamefont {P.}~\bibnamefont
  {T\"{o}rm\"{a}}},\ }\href@noop {} {\bibinfo {title} {Fermi surface
  deformations and pairing in mixtures of dipolar and non-dipolar fermions}}
  (\bibinfo {year} {2017}),\ \Eprint {https://arxiv.org/abs/1612.07953}
  {arXiv:1612.07953 [cond-mat.quant-gas]} \BibitemShut {NoStop}%
\bibitem [{\citenamefont {Nishimura}\ \emph {et~al.}(2021)\citenamefont
  {Nishimura}, \citenamefont {Nakano}, \citenamefont {Iida}, \citenamefont
  {Tajima}, \citenamefont {Miyakawa},\ and\ \citenamefont
  {Yabu}}]{Nishimura_2021}%
  \BibitemOpen
  \bibfield  {author} {\bibinfo {author} {\bibfnamefont {K.}~\bibnamefont
  {Nishimura}}, \bibinfo {author} {\bibfnamefont {E.}~\bibnamefont {Nakano}},
  \bibinfo {author} {\bibfnamefont {K.}~\bibnamefont {Iida}}, \bibinfo {author}
  {\bibfnamefont {H.}~\bibnamefont {Tajima}}, \bibinfo {author} {\bibfnamefont
  {T.}~\bibnamefont {Miyakawa}},\ and\ \bibinfo {author} {\bibfnamefont
  {H.}~\bibnamefont {Yabu}},\ }\bibfield  {title} {\bibinfo {title} {Ground
  state of the polaron in an ultracold dipolar fermi gas},\ }\href
  {https://doi.org/10.1103/physreva.103.033324} {\bibfield  {journal} {\bibinfo
   {journal} {Phys.~Rev.~A}\ }\textbf {\bibinfo {volume} {103}},\ \bibinfo
  {pages} {033324} (\bibinfo {year} {2021})}\BibitemShut {NoStop}%
\bibitem [{\citenamefont {Chevy}\ and\ \citenamefont
  {Mora}(2010)}]{Chevy_2010}%
  \BibitemOpen
  \bibfield  {author} {\bibinfo {author} {\bibfnamefont {F.}~\bibnamefont
  {Chevy}}\ and\ \bibinfo {author} {\bibfnamefont {C.}~\bibnamefont {Mora}},\
  }\bibfield  {title} {\bibinfo {title} {Ultra-cold polarized fermi gases},\
  }\href {https://doi.org/10.1088/0034-4885/73/11/112401} {\bibfield  {journal}
  {\bibinfo  {journal} {Reports on Progress in Physics}\ }\textbf {\bibinfo
  {volume} {73}},\ \bibinfo {pages} {112401} (\bibinfo {year}
  {2010})}\BibitemShut {NoStop}%
\bibitem [{\citenamefont {Massignan}\ \emph {et~al.}(2014)\citenamefont
  {Massignan}, \citenamefont {Zaccanti},\ and\ \citenamefont
  {Bruun}}]{Massignan_2014}%
  \BibitemOpen
  \bibfield  {author} {\bibinfo {author} {\bibfnamefont {P.}~\bibnamefont
  {Massignan}}, \bibinfo {author} {\bibfnamefont {M.}~\bibnamefont
  {Zaccanti}},\ and\ \bibinfo {author} {\bibfnamefont {G.~M.}\ \bibnamefont
  {Bruun}},\ }\bibfield  {title} {\bibinfo {title} {Polarons, dressed molecules
  and itinerant ferromagnetism in ultracold fermi gases},\ }\href
  {https://doi.org/10.1088/0034-4885/77/3/034401} {\bibfield  {journal}
  {\bibinfo  {journal} {Reports on Progress in Physics}\ }\textbf {\bibinfo
  {volume} {77}},\ \bibinfo {pages} {034401} (\bibinfo {year}
  {2014})}\BibitemShut {NoStop}%
\bibitem [{\citenamefont {Schmidt}\ \emph {et~al.}(2018)\citenamefont
  {Schmidt}, \citenamefont {Knap}, \citenamefont {Ivanov}, \citenamefont {You},
  \citenamefont {Cetina},\ and\ \citenamefont {Demler}}]{Schmidt_2018}%
  \BibitemOpen
  \bibfield  {author} {\bibinfo {author} {\bibfnamefont {R.}~\bibnamefont
  {Schmidt}}, \bibinfo {author} {\bibfnamefont {M.}~\bibnamefont {Knap}},
  \bibinfo {author} {\bibfnamefont {D.~A.}\ \bibnamefont {Ivanov}}, \bibinfo
  {author} {\bibfnamefont {J.-S.}\ \bibnamefont {You}}, \bibinfo {author}
  {\bibfnamefont {M.}~\bibnamefont {Cetina}},\ and\ \bibinfo {author}
  {\bibfnamefont {E.}~\bibnamefont {Demler}},\ }\bibfield  {title} {\bibinfo
  {title} {Universal many-body response of heavy impurities coupled to a fermi
  sea: a review of recent progress},\ }\href
  {https://doi.org/10.1088/1361-6633/aa9593} {\bibfield  {journal} {\bibinfo
  {journal} {Reports on Progress in Physics}\ }\textbf {\bibinfo {volume}
  {81}},\ \bibinfo {pages} {024401} (\bibinfo {year} {2018})}\BibitemShut
  {NoStop}%
\bibitem [{\citenamefont {Tajima}\ \emph {et~al.}(2021)\citenamefont {Tajima},
  \citenamefont {Takahashi}, \citenamefont {Mistakidis}, \citenamefont
  {Nakano},\ and\ \citenamefont {Iida}}]{Tajima_2021}%
  \BibitemOpen
  \bibfield  {author} {\bibinfo {author} {\bibfnamefont {H.}~\bibnamefont
  {Tajima}}, \bibinfo {author} {\bibfnamefont {J.}~\bibnamefont {Takahashi}},
  \bibinfo {author} {\bibfnamefont {S.}~\bibnamefont {Mistakidis}}, \bibinfo
  {author} {\bibfnamefont {E.}~\bibnamefont {Nakano}},\ and\ \bibinfo {author}
  {\bibfnamefont {K.}~\bibnamefont {Iida}},\ }\bibfield  {title} {\bibinfo
  {title} {Polaron problems in ultracold atoms: Role of a fermi sea across
  different spatial dimensions and quantum fluctuations of a bose medium},\
  }\href {https://doi.org/10.3390/atoms9010018} {\bibfield  {journal} {\bibinfo
   {journal} {Atoms}\ }\textbf {\bibinfo {volume} {9}},\ \bibinfo {pages} {18}
  (\bibinfo {year} {2021})}\BibitemShut {NoStop}%
\bibitem [{\citenamefont {Scazza}\ \emph {et~al.}(2022)\citenamefont {Scazza},
  \citenamefont {Zaccanti}, \citenamefont {Massignan}, \citenamefont {Parish},\
  and\ \citenamefont {Levinsen}}]{Scazza_2022}%
  \BibitemOpen
  \bibfield  {author} {\bibinfo {author} {\bibfnamefont {F.}~\bibnamefont
  {Scazza}}, \bibinfo {author} {\bibfnamefont {M.}~\bibnamefont {Zaccanti}},
  \bibinfo {author} {\bibfnamefont {P.}~\bibnamefont {Massignan}}, \bibinfo
  {author} {\bibfnamefont {M.~M.}\ \bibnamefont {Parish}},\ and\ \bibinfo
  {author} {\bibfnamefont {J.}~\bibnamefont {Levinsen}},\ }\bibfield  {title}
  {\bibinfo {title} {Repulsive fermi and bose polarons in quantum gases},\
  }\href {https://doi.org/10.3390/atoms10020055} {\bibfield  {journal}
  {\bibinfo  {journal} {Atoms}\ }\textbf {\bibinfo {volume} {10}},\ \bibinfo
  {pages} {55} (\bibinfo {year} {2022})}\BibitemShut {NoStop}%
\bibitem [{\citenamefont {Camacho-Guardian}\ \emph {et~al.}(2018)\citenamefont
  {Camacho-Guardian}, \citenamefont {Ardila}, \citenamefont {Pohl},\ and\
  \citenamefont {Bruun}}]{Camacho_Guardian_2018}%
  \BibitemOpen
  \bibfield  {author} {\bibinfo {author} {\bibfnamefont {A.}~\bibnamefont
  {Camacho-Guardian}}, \bibinfo {author} {\bibfnamefont {L.~A.~P.}\
  \bibnamefont {Ardila}}, \bibinfo {author} {\bibfnamefont {T.}~\bibnamefont
  {Pohl}},\ and\ \bibinfo {author} {\bibfnamefont {G.~M.}\ \bibnamefont
  {Bruun}},\ }\bibfield  {title} {\bibinfo {title} {Bipolarons in a
  bose-einstein condensate},\ }\href
  {https://doi.org/10.1103/physrevlett.121.013401} {\bibfield  {journal}
  {\bibinfo  {journal} {Phys.~Rev.~Lett.}\ }\textbf {\bibinfo {volume} {121}},\
  \bibinfo {pages} {013401} (\bibinfo {year} {2018})}\BibitemShut {NoStop}%
\bibitem [{\citenamefont {Moriya}\ \emph {et~al.}(2021)\citenamefont {Moriya},
  \citenamefont {Tajima}, \citenamefont {Horiuchi}, \citenamefont {Iida},\ and\
  \citenamefont {Nakano}}]{Moriya_2021}%
  \BibitemOpen
  \bibfield  {author} {\bibinfo {author} {\bibfnamefont {H.}~\bibnamefont
  {Moriya}}, \bibinfo {author} {\bibfnamefont {H.}~\bibnamefont {Tajima}},
  \bibinfo {author} {\bibfnamefont {W.}~\bibnamefont {Horiuchi}}, \bibinfo
  {author} {\bibfnamefont {K.}~\bibnamefont {Iida}},\ and\ \bibinfo {author}
  {\bibfnamefont {E.}~\bibnamefont {Nakano}},\ }\bibfield  {title} {\bibinfo
  {title} {Binding two and three alpha particles in cold neutron matter},\
  }\href {https://doi.org/10.1103/physrevc.104.065801} {\bibfield  {journal}
  {\bibinfo  {journal} {Phys.~Rev.~C}\ }\textbf {\bibinfo {volume} {104}},\
  \bibinfo {pages} {065801} (\bibinfo {year} {2021})}\BibitemShut {NoStop}%
\bibitem [{\citenamefont {Tajima}\ \emph {et~al.}(2022)\citenamefont {Tajima},
  \citenamefont {Moriya}, \citenamefont {Horiuchi}, \citenamefont {Iida},\ and\
  \citenamefont {Nakano}}]{Tajima_2022}%
  \BibitemOpen
  \bibfield  {author} {\bibinfo {author} {\bibfnamefont {H.}~\bibnamefont
  {Tajima}}, \bibinfo {author} {\bibfnamefont {H.}~\bibnamefont {Moriya}},
  \bibinfo {author} {\bibfnamefont {W.}~\bibnamefont {Horiuchi}}, \bibinfo
  {author} {\bibfnamefont {K.}~\bibnamefont {Iida}},\ and\ \bibinfo {author}
  {\bibfnamefont {E.}~\bibnamefont {Nakano}},\ }\bibfield  {title} {\bibinfo
  {title} {Resonance-to-bound transition of in neutron matter and its analogy
  with heteronuclear feshbach molecules},\ }\href
  {https://doi.org/10.1103/physrevc.106.045807} {\bibfield  {journal} {\bibinfo
   {journal} {Phys.~Rev.~C}\ }\textbf {\bibinfo {volume} {106}},\ \bibinfo
  {pages} {045807} (\bibinfo {year} {2022})}\BibitemShut {NoStop}%
\bibitem [{\citenamefont {Muir}\ \emph {et~al.}(2022)\citenamefont {Muir},
  \citenamefont {Levinsen}, \citenamefont {Earl}, \citenamefont {Conway},
  \citenamefont {Cole}, \citenamefont {Wurdack}, \citenamefont {Mishra},
  \citenamefont {Ing}, \citenamefont {Estrecho}, \citenamefont {Lu},
  \citenamefont {Efimkin}, \citenamefont {Tollerud}, \citenamefont
  {Ostrovskaya}, \citenamefont {Parish},\ and\ \citenamefont
  {Davis}}]{Muir_2022}%
  \BibitemOpen
  \bibfield  {author} {\bibinfo {author} {\bibfnamefont {J.~B.}\ \bibnamefont
  {Muir}}, \bibinfo {author} {\bibfnamefont {J.}~\bibnamefont {Levinsen}},
  \bibinfo {author} {\bibfnamefont {S.~K.}\ \bibnamefont {Earl}}, \bibinfo
  {author} {\bibfnamefont {M.~A.}\ \bibnamefont {Conway}}, \bibinfo {author}
  {\bibfnamefont {J.~H.}\ \bibnamefont {Cole}}, \bibinfo {author}
  {\bibfnamefont {M.}~\bibnamefont {Wurdack}}, \bibinfo {author} {\bibfnamefont
  {R.}~\bibnamefont {Mishra}}, \bibinfo {author} {\bibfnamefont {D.~J.}\
  \bibnamefont {Ing}}, \bibinfo {author} {\bibfnamefont {E.}~\bibnamefont
  {Estrecho}}, \bibinfo {author} {\bibfnamefont {Y.}~\bibnamefont {Lu}},
  \bibinfo {author} {\bibfnamefont {D.~K.}\ \bibnamefont {Efimkin}}, \bibinfo
  {author} {\bibfnamefont {J.~O.}\ \bibnamefont {Tollerud}}, \bibinfo {author}
  {\bibfnamefont {E.~A.}\ \bibnamefont {Ostrovskaya}}, \bibinfo {author}
  {\bibfnamefont {M.~M.}\ \bibnamefont {Parish}},\ and\ \bibinfo {author}
  {\bibfnamefont {J.~A.}\ \bibnamefont {Davis}},\ }\bibfield  {title} {\bibinfo
  {title} {Interactions between fermi polarons in monolayer {WS}2},\ }\href
  {https://doi.org/10.1038/s41467-022-33811-x} {\bibfield  {journal} {\bibinfo
  {journal} {Nature Communications}\ }\textbf {\bibinfo {volume} {13}},\
  \bibinfo {pages} {6164} (\bibinfo {year} {2022})}\BibitemShut {NoStop}%
\bibitem [{\citenamefont {Ruderman}\ and\ \citenamefont
  {Kittel}(1954)}]{Ruderman_1954}%
  \BibitemOpen
  \bibfield  {author} {\bibinfo {author} {\bibfnamefont {M.}~\bibnamefont
  {Ruderman}}\ and\ \bibinfo {author} {\bibfnamefont {C.}~\bibnamefont
  {Kittel}},\ }\bibfield  {title} {\bibinfo {title} {Indirect exchange coupling
  of nuclear magnetic moments by conduction electrons},\ }\href
  {https://doi.org/https://doi.org/10.1103/PhysRev.96.99} {\bibfield  {journal}
  {\bibinfo  {journal} {Physical Review}\ }\textbf {\bibinfo {volume} {96}},\
  \bibinfo {pages} {99} (\bibinfo {year} {1954})}\BibitemShut {NoStop}%
\bibitem [{\citenamefont {Kasuya}(1956)}]{Kasuya_1956}%
  \BibitemOpen
  \bibfield  {author} {\bibinfo {author} {\bibfnamefont {T.}~\bibnamefont
  {Kasuya}},\ }\bibfield  {title} {\bibinfo {title} {A theory of metallic
  ferro- and antiferromagnetism on zener's model},\ }\href
  {https://doi.org/https://doi.org/10.1143/PTP.16.45} {\bibfield  {journal}
  {\bibinfo  {journal} {Progress of Theoretical Physics}\ }\textbf {\bibinfo
  {volume} {16}},\ \bibinfo {pages} {45} (\bibinfo {year} {1956})}\BibitemShut
  {NoStop}%
\bibitem [{\citenamefont {Yoshida}(1957)}]{Yoshida_1957}%
  \BibitemOpen
  \bibfield  {author} {\bibinfo {author} {\bibfnamefont {K.}~\bibnamefont
  {Yoshida}},\ }\bibfield  {title} {\bibinfo {title} {Magnetic properties of
  cu-mn alloys},\ }\href
  {https://doi.org/https://doi.org/10.1103/PhysRev.106.893} {\bibfield
  {journal} {\bibinfo  {journal} {Physical Review}\ }\textbf {\bibinfo {volume}
  {106}},\ \bibinfo {pages} {893} (\bibinfo {year} {1957})}\BibitemShut
  {NoStop}%
\bibitem [{\citenamefont {Fetter}\ and\ \citenamefont
  {Walecka}(2003)}]{FW_2003}%
  \BibitemOpen
  \bibfield  {author} {\bibinfo {author} {\bibfnamefont {A.~L.}\ \bibnamefont
  {Fetter}}\ and\ \bibinfo {author} {\bibfnamefont {J.~D.}\ \bibnamefont
  {Walecka}},\ }\href@noop {} {\emph {\bibinfo {title} {Quantum Theory of
  Many-Particle Systems}}}\ (\bibinfo  {publisher} {Dover Publication},\
  \bibinfo {year} {2003})\BibitemShut {NoStop}%
\bibitem [{\citenamefont {Kashimura}\ \emph {et~al.}(2012)\citenamefont
  {Kashimura}, \citenamefont {Watanabe},\ and\ \citenamefont
  {Ohashi}}]{Kashimura_2012}%
  \BibitemOpen
  \bibfield  {author} {\bibinfo {author} {\bibfnamefont {T.}~\bibnamefont
  {Kashimura}}, \bibinfo {author} {\bibfnamefont {R.}~\bibnamefont
  {Watanabe}},\ and\ \bibinfo {author} {\bibfnamefont {Y.}~\bibnamefont
  {Ohashi}},\ }\bibfield  {title} {\bibinfo {title} {Spin susceptibility and
  fluctuation corrections in the {BCS}-{BEC} crossover regime of an ultracold
  fermi gas},\ }\href {https://doi.org/10.1103/physreva.86.043622} {\bibfield
  {journal} {\bibinfo  {journal} {Phys.~Rev.~A}\ }\textbf {\bibinfo {volume}
  {86}},\ \bibinfo {pages} {043622} (\bibinfo {year} {2012})}\BibitemShut
  {NoStop}%
\bibitem [{\citenamefont {Tajima}\ and\ \citenamefont
  {Uchino}(2018)}]{Tajima_2018}%
  \BibitemOpen
  \bibfield  {author} {\bibinfo {author} {\bibfnamefont {H.}~\bibnamefont
  {Tajima}}\ and\ \bibinfo {author} {\bibfnamefont {S.}~\bibnamefont
  {Uchino}},\ }\bibfield  {title} {\bibinfo {title} {Many fermi polarons at
  nonzero temperature},\ }\href {https://doi.org/10.1088/1367-2630/aad1e7}
  {\bibfield  {journal} {\bibinfo  {journal} {New Journal of Physics}\ }\textbf
  {\bibinfo {volume} {20}},\ \bibinfo {pages} {073048} (\bibinfo {year}
  {2018})}\BibitemShut {NoStop}%
\bibitem [{\citenamefont {Ring}\ and\ \citenamefont {Shuck}(2004)}]{RS_2004}%
  \BibitemOpen
  \bibfield  {author} {\bibinfo {author} {\bibfnamefont {P.}~\bibnamefont
  {Ring}}\ and\ \bibinfo {author} {\bibfnamefont {P.}~\bibnamefont {Shuck}},\
  }\href@noop {} {\emph {\bibinfo {title} {The Nuclear Many-Body Problem -- 3rd
  edition}}}\ (\bibinfo  {publisher} {Springer},\ \bibinfo {year}
  {2004})\BibitemShut {NoStop}%
\end{thebibliography}%

\appendix 
\section{Polarization function at static limit} 
\label{app:A}
The density-density correlation function of dipolar fermions to the single-loop order 
reduces to the polarization function. 
Its static limit is calculated as follows: 
\beq
\Pi_{0}(\bq) &=& 
\sum_\bk \frac{f_\bk - f_{\bk+\bq}}{\epsilon_\bk - \epsilon_{\bk+\bq}}
\nn
&=&2\sum_\bk \frac{f_\bk}{\epsilon_\bk - \epsilon_{\bk+\bq}}
\nn
&=& 
\frac{4m_1}{(2\pi)^3\lambda^2}
\int_{-k_F/\beta }^{k_F/\beta }{\rm d}k_z 
\int_{k_t \le \sqrt{\beta k_F^2-\beta^3k_z^2}} {\rm d}^2\bk_t\, 
\nn &&\times
\frac{1}{
\beta^{-1} \bk_t^2 + \beta^2 k_z^2 - \beta^{-1}\lk \bk_t-\bq_t\rk^2 -\beta^2 \lk k_z-q_z\rk^2}, 
\nn
\eeq
with $\bk_t=\lk k_x, k_y \rk$ and $\bq_t=\lk q_x, q_y \rk$ being the momenta 
projected in the transverse plane. 
Here changing of the variables as 
$\lk \bk_t, k_z\rk =\lk \beta^{1/2}\tilde{\bk}_t, \beta^{-1}\tilde{k}_z \rk k_F$ and 
$\lk \bq_t, q_z\rk =\lk \beta^{1/2}\tilde{\bq}_t, \beta^{-1}\tilde{q}_z \rk k_F$ 
in order for the spherical symmetry to be restored in the integral, 
we obtain 
\beq
\Pi_{0}(\bq) 
&=&
\frac{m_1 k_F}{2\pi^3\lambda^2}
\int_{|\tilde{\bk}|\le 1} {\rm d}^3\tilde{\bk}\, 
\frac{1}{\tilde{\bk}^2 -\lk \tilde{\bk}-\tilde{\bq}\rk^2 } 
\nn
\nn
&=& 
\frac{m_1 k_F}{2\tilde{q}\pi^2\lambda^2}
\int_{0}^{1}{\rm d}\tilde{k} \tilde{k}
\int_{-1}^{1}{\rm d}x \, \frac{1}{x-\tilde{q}/2\tilde{k}}
\nn
&=& \frac{m_1k_F}{2q\pi^2\lambda^2}
\int_{0}^{1}{\rm d}\tilde{k} \tilde{k}\, 
\log\left|\frac{\tilde{k}-\tilde{q}/2}{\tilde{k}+\tilde{q}/2}\right|
\nn
&=&
-\frac{m_1k_F}{4\pi^2\lambda^2}
\ldk 
1+
\frac{4-\tilde{q}^2}{4\tilde{q}}\log\left|\frac{1+\tilde{q}/2}{1-\tilde{q}/2}\right|
\rdk 
\eeq
where $\tilde{q}=|\tilde{\bq}|$ 
with $\tilde{\bq}=k_F^{-1} \lk \beta^{-1/2} \bq_t, \beta q_z \rk$.  

\section{Induced interaction potential}
\label{app:B}
We calculate the induced interaction potential 
between two probe impurities 
using the static density-density correlation function 
to the single-loop order as follows: 
\beq
V(\bx) &=& g^2\sum_\bq \, e^{i \bx\cdot \bq}\, \Pi_0(|\tilde{\bq}_\beta|) 
\nn 
&=& -g^2\frac{m_1k_F}{4\pi^2\lambda^2} \int \frac{{\rm d}^3\bq}{(2\pi)^3} \, 
e^{i \bx\cdot \bq}
\ldk 
1 + \frac{4-\tilde{q}^2}{4 \tilde{q}}\ln\left|\frac{\tilde{q}+2}{\tilde{q}-2}\right|
\rdk
\nn 
&=& -g^2\frac{m_1 k_F^4}{4\pi^2\lambda^2} \int \frac{{\rm d}^3\tilde{\bq}}{(2\pi)^3} \, 
e^{i \tilde{\br}\cdot \tilde{\bq}}
\ldk 
1 + \frac{4-\tilde{q}^2}{4 \tilde{q}}\ln\left|\frac{\tilde{q}+2}{\tilde{q}-2}\right|
\rdk
\nn
&=& 
 -g^2\frac{m_1k_F^4}{16\pi^4\lambda^2 i\tilde{r}} 
 \int_{-\infty}^\infty {\rd}\tilde{q}  \, 
e^{i\tilde{r}\tilde{q}}
\tilde{q}
\ldk 
1 + \frac{4-\tilde{q}^2}{4 \tilde{q}}\ln\left|\frac{\tilde{q}+2}{\tilde{q}-2}\right|
\rdk
\label{app1}
\nn 
\eeq
where the anisotropy has been taken over by the coordinate vector:
\beq
\tilde{\br}&= &\lk \beta^{1/2}x, \beta^{1/2}y, \beta^{-1} z\rk k_F, 
\\ 
\tilde{r}&=& \sqrt{\beta \lk x^2 + y^2\rk + \beta^{-2} z^2} k_F. 
\eeq
Hereafter we will change the notation as $\tilde{r}\rightarrow r$ and $\tilde{q}\rightarrow q$ 
for notational simplicity for a while. 
In the last line of (\ref{app1}) 
the integrand has branch points at $q=\pm2$, 
thus we can make a cut between them. 
\begin{figure}[h]
  \begin{center}
 \resizebox{85mm}{!}{\includegraphics{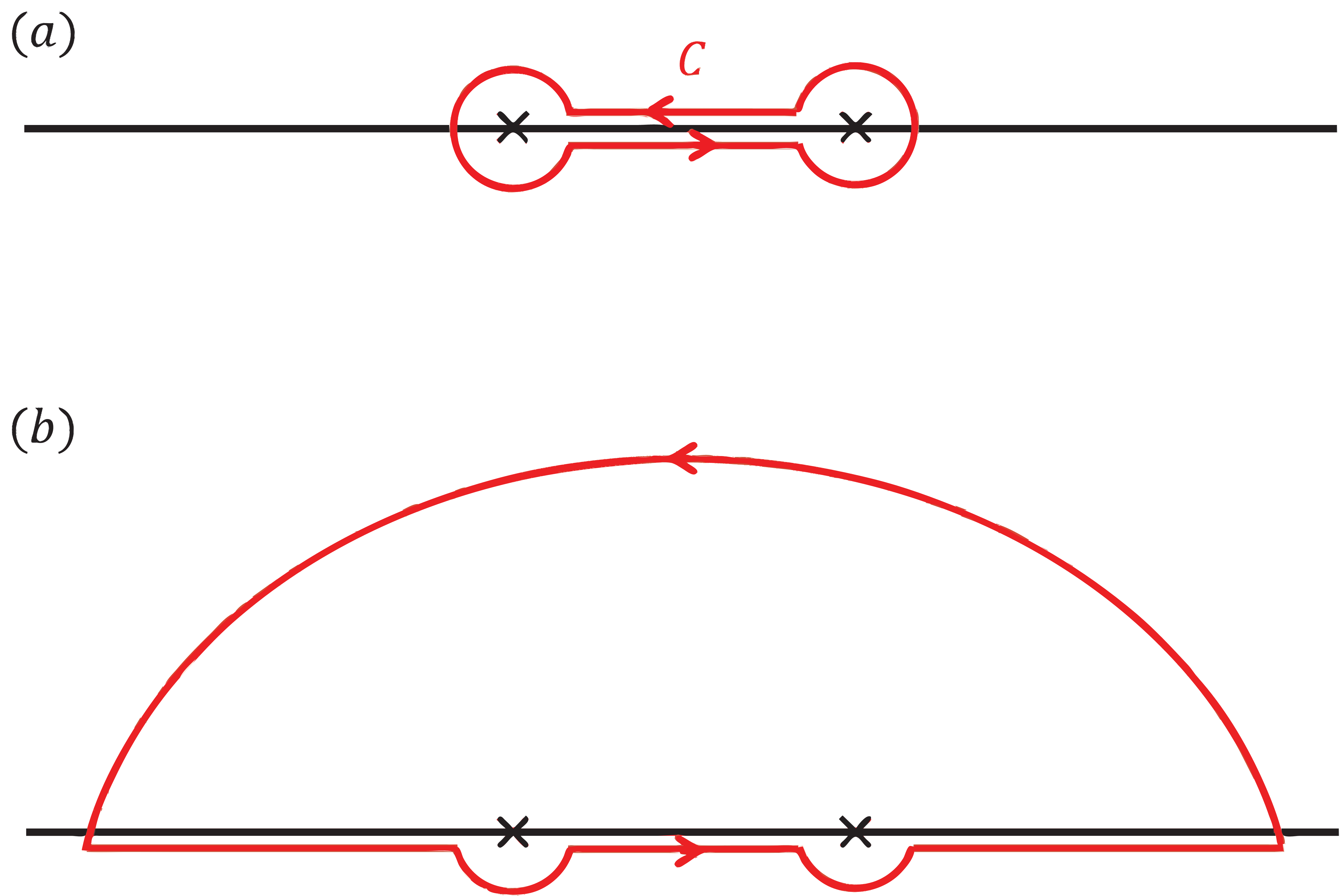}} 
    \caption{(a) Integral contour $C$ in complex $q$ plane. Cross symbols represent the branch points $q=\pm2$. 
(b) Deformation of contour $C$. 
The path in real axis is extended to $q \rightarrow \pm \infty$. }
    \label{comp1}
  \end{center}
\end{figure}
Now we consider that loop integral $I$ defined just below, along the path $C$ enclosing the branch cut as depicted in Fig~\ref{comp1}(a), 
which is calculated with analytic functions defined around the cut as 
\beq
I&\equiv&\int_C {\rm d}q \, e^{irq}
\ldk 
q + \frac{4-q^2}{8}\ln\lk \frac{q+2}{q-2}\rk^2
\rdk
\nn
&=&
\int_{2-\epsilon}^{-2+\epsilon}{\rm d}s\, 
e^{irs}\frac{4-s^2}{4}\ln\frac{s+2}{(2-s)e^{\pi i}}
\nn&&
+ 
\int_{-2+\epsilon}^{2-\epsilon}{\rm d}s\, 
e^{irs}\frac{4-s^2}{4}\ln\frac{(s+2)e^{2\pi i}}{(2-s)e^{-\pi i}}
\nn
&=&
4\pi i \int_{-2}^{2}{\rm d}s\, 
e^{irs}\frac{4-s^2}{4}. 
\eeq

On the other hand, the integral $I$ has an another form, 
that is, $I$ is equivalent to the integral along the real axis since the contribution 
from the upper hemisphere vanishes as shown in Fig~\ref{comp1}(b). 
Using the same analytic functions around the cut used above, 
we obtain 
\beq
I &=& 
\int_{-\infty}^\infty {\rm d}q \, e^{irq}
\ldk 
q + \frac{4-q^2}{8}\ln\lk \frac{q+2}{q-2}\rk^2
\rdk
\nn
&=& 
\lk \int_{-\infty}^{-2-\epsilon} +
\int_{2+\epsilon}^\infty \rk {\rm d}q \, e^{irq}
\ldk 
q + \frac{4-q^2}{4}\ln\lk \frac{q+2}{q-2}\rk
\rdk
\nn
&&+ 
\int_{-2+\epsilon}^{2-\epsilon} {\rm d}q \, e^{irq}
\ldk 
q + \frac{4-q^2}{4}\ln\frac{(q+2)e^{2\pi i}}{(2-q)e^{-\pi i}}
\rdk
\nn
&=& 
\int_{-\infty}^\infty {\rm d}q \, e^{irq}
\ldk 
q + \frac{4-q^2}{4}\ln\left|\frac{q+2}{q-2}\right|
\rdk
\nn&&\quad +
3\pi i \int_{-2}^{2} {\rm d}q \, e^{irq}
\frac{4-q^2}{4}. 
\eeq
Comparing the two expressions of $I$, 
we find 
\beq
&&\int_{-\infty}^\infty {\rm d}q \, e^{irq}
\ldk 
q + \frac{4-q^2}{4}\ln\left|\frac{q+2}{q-2}\right|
\rdk
\nn
&=& \pi i \int_{-2}^{2} {\rm d}q \, e^{irq}
\frac{4-q^2}{4}
\nn
&=& 
\frac{\pi i}{2} 
\lk \frac{2\sin2r}{r^3}  -\frac{4\cos2r}{r^2}  \rk. 
\eeq
Getting all together, after recovering the original notation of $\tilde{\br}$ 
we obtain
\beq
V(\bx) &=& 
 g^2\frac{m_1 k_F^4}{16\pi^3\lambda^2} 
 \lk \frac{2\cos2\tilde{r}}{\tilde{r}^3} -\frac{\sin2\tilde{r}}{\tilde{r}^4}  \rk. 
\label{rkky1}
\eeq
This is nothing but the RKKY potential except for the anisotropy in $\tilde{r}$ 
for $\beta\neq1$. 

\section{Critical coupling strength in spherically symmetric case}
\label{app:D}
\begin{figure}[h]
  \begin{tabular}{c}
 \resizebox{85mm}{!}{\includegraphics{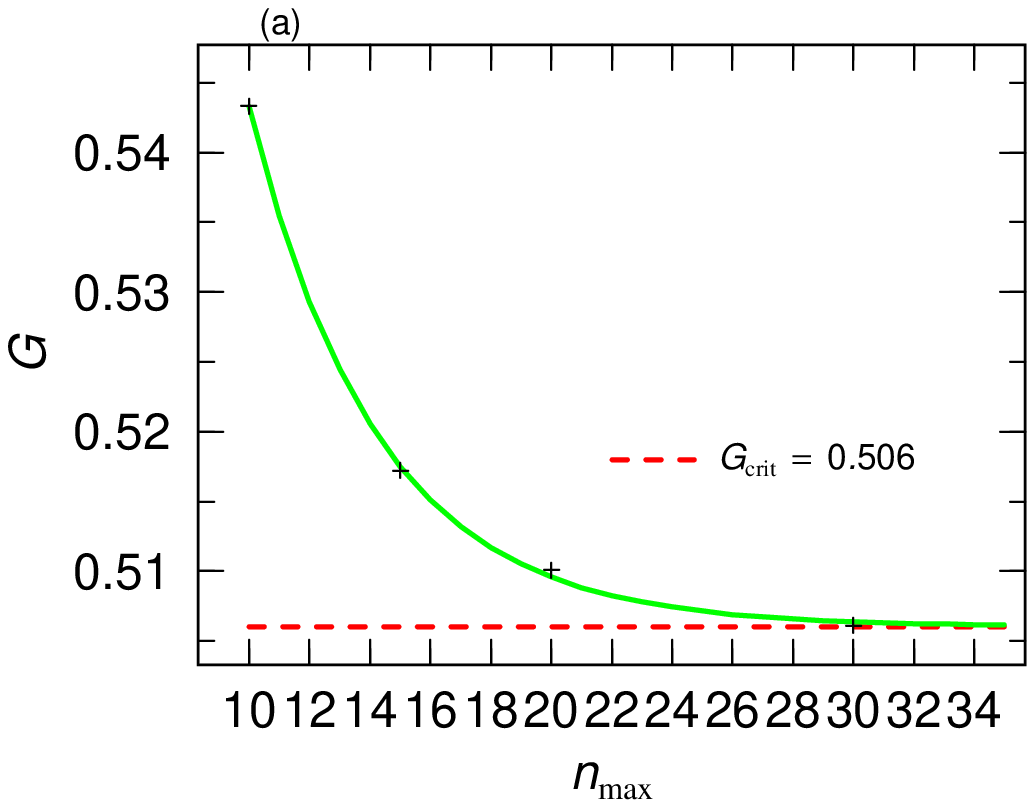}} \\
 \resizebox{85mm}{!}{\includegraphics{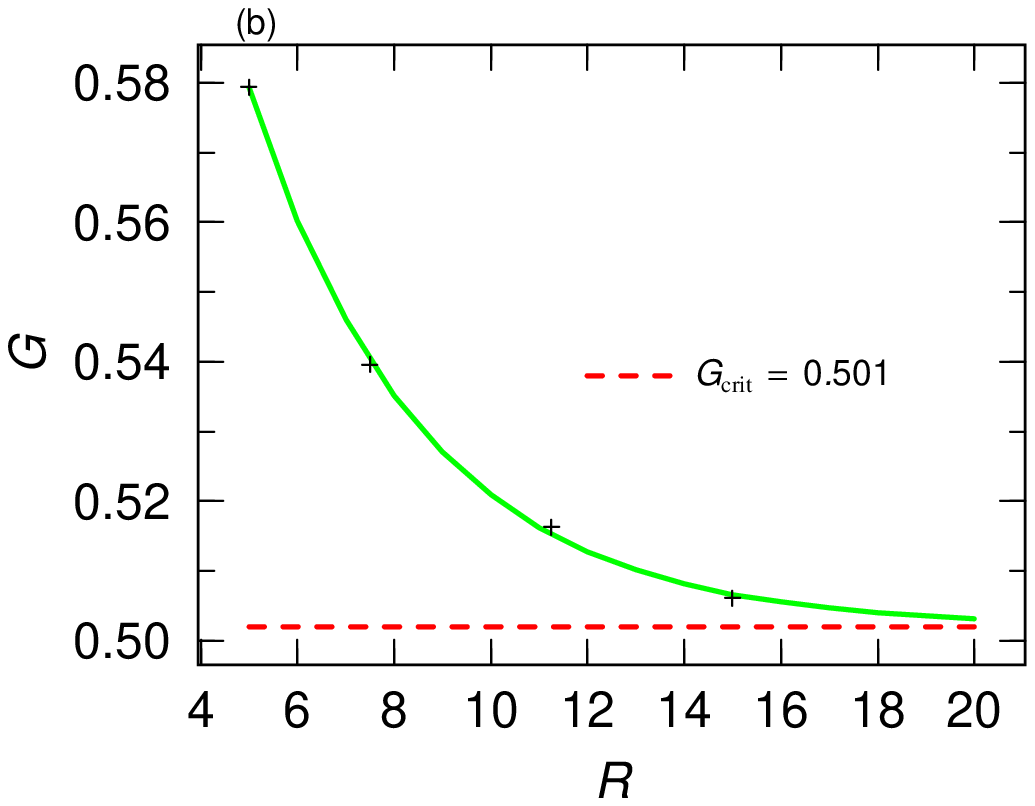}} 
    \end{tabular}
    \caption{(a) Numerical values of critical coupling strength $G$ 
as a function of $i_{\rm max}$ in the case of $R=15$. 
Cross symbols correspond to $i_{\rm max}=10,15,20,30$, respectively. 
Solid line corresponds to the function (\ref{modelA}), in which parameters $a,b,c$ are determined 
by the $\chi$-square fitting of those symbols; 
Dashed line corresponds to the parameter $a=G_{\rm crit}$,  
a critical coupling strength deduced for $n_{\rm max}\rightarrow \infty$. 
(b) Numerical values of critical coupling strength $G$ 
as a function of $R$ in the case of  $i_{\rm max}=30$. 
Symbols correspond to $R=5, 7.5, 11.25, 15$, respectively. 
Solid line is the $\chi$-square fitting of these symbols using the function (\ref{modelA}).  
Dashed line corresponds to the parameter $a=G_{\rm crit}$:  
a deduced critical coupling strength at $R\rightarrow \infty$
i.e, in the spatially uniform system.}
    \label{figLnsph}
\end{figure}

In this appendix, we present the numerical results for the spherically symmetric RKKY potential 
in the polar coordinates to confront with the case of $\beta=1$ in the cylindrical coordinates. 
In this case the Schr\"{o}dinger equation for the radial coordinate $r$, corresponding to (\ref{scheq1}), becomes 
\begin{equation}
0= \ldk \frac{1}{r^2}\partial_r \lk \frac{1}{r^2} \partial_r \rk 
+\frac{L(L+1)}{r^2} - v(r) +\varepsilon \rdk \psi_{L}(r), 
\label{schesphq1} 
\end{equation}
where $L=0,1,2,\cdots$ denotes the angular momentum. 
Now we expand the wave function as 
\beq
\psi_{L}(r) &=& \sum_{i=1,2,3,\cdots}  f_{L}(i)\,  j_{{L};i}(r) 
 \label{funcsph2}
\eeq
in terms of the spherical Bessel function of the first kind $j_{L}(r)$: 
\beq
j_{{L};i}(r) &:=&\frac{1}{\sqrt{{\mathcal N}_i}} j_{L}(s_i r/R)
\eeq
with 
${\mathcal N}_i \equiv -\pi R^3\frac{J_{L-\frac{1}{2}}(s_i) J_{L+\frac{3}{2}}(s_i)}{4 s_i}$ 
and $s_i$ being the zero's of the Bessel function $J_{L+\frac{1}{2}}(s_i)=0$,  
so that it satisfies the normalization condition 
\beq
\int_{0}^{R} {\rm d}r r^2\,  j_{{L};i}(r)  j_{{L};j}(r) = \delta_{ij}. 
\eeq
Truncation of the number of the spherical Bessel functions within $i\le i_{\rm max}$ 
leads to the matrix eigenvalue equation 
\beq
\sum_{j=1,2,\cdots, i_{\rm max}} \ldk \lk \frac{s_i}{R}\rk^2 \delta_{ij} + v_{ij} -\epsilon \delta_{ij} \rdk f_L(j)=0
\eeq
where 
$
v_{ij} \equiv 
\int_0^R  {\rm d}r \, r^2\, j_{{L};i}(r)\, v(r)\, j_{{L};j}(r). 
$
Solving the matrix equation, 
we show the numerical result of the critical coupling strength for the first bound state in Fig.~\ref{figLnsph} 
to examine its dependence on (a) the number of basis functions $i_{\rm max}$ and (b) the system size $R$. 
In order to estimate the critical value in the spatially uniform system, 
we make the extrapolation procedure explained in Appendix~\ref{app:C} in detail, and 
find that $G_{\rm crit}=0.501$ at $R\rightarrow \infty$.  

\section{Critical coupling strength for bound state}
\label{app:C}
Here we examine the finite size effect on numerical results, and 
estimate a critical value of the dimensionless coupling strength (\ref{dimensionlessG1}), 
above which the first bound state emerges in the spatially uniform system. 
In the wave function expansion, 
we denote the maximum quantum number of a truncated set of the plane wave functions (\ref{func1}) 
by $n_{\rm max}$ and that of the Bessel functions (\ref{func2}) by $i_{\rm max}$, 
and consider some cases where 
$n_{\rm max}=i_{\rm max}=15, 20, 30$ and $L=2R=10, 15, 22.5, 30$ for the sizes of cylinder in the unit of $k_F^{-1}$. 
\begin{itemize}
\item
In Fig.~\ref{figLn}(a) 
we first show 
the numerical result for the critical coupling strength $G$
as a function of $n_{\rm max}$ (symbols) for $L=2R=30$, 
together with the result from the fitting function given by 
\beq
G_{\rm fit}(x)&=&a + b\, e^{-c x}. 
\label{modelA}
\eeq
We determine parameters $a,b,c$ using $\chi$-square fitting to the symbols, 
and estimate a critical coupling strength $G_{\rm crit}=a$ at $n_{\rm max}\rightarrow \infty$. 
From these results we expect that $n_{\rm max}=i_{\rm max}=30$ already gives a good convergence.  
\begin{figure}[h]
  \begin{center}
  \begin{tabular}{c}
 \resizebox{85mm}{!}{\includegraphics{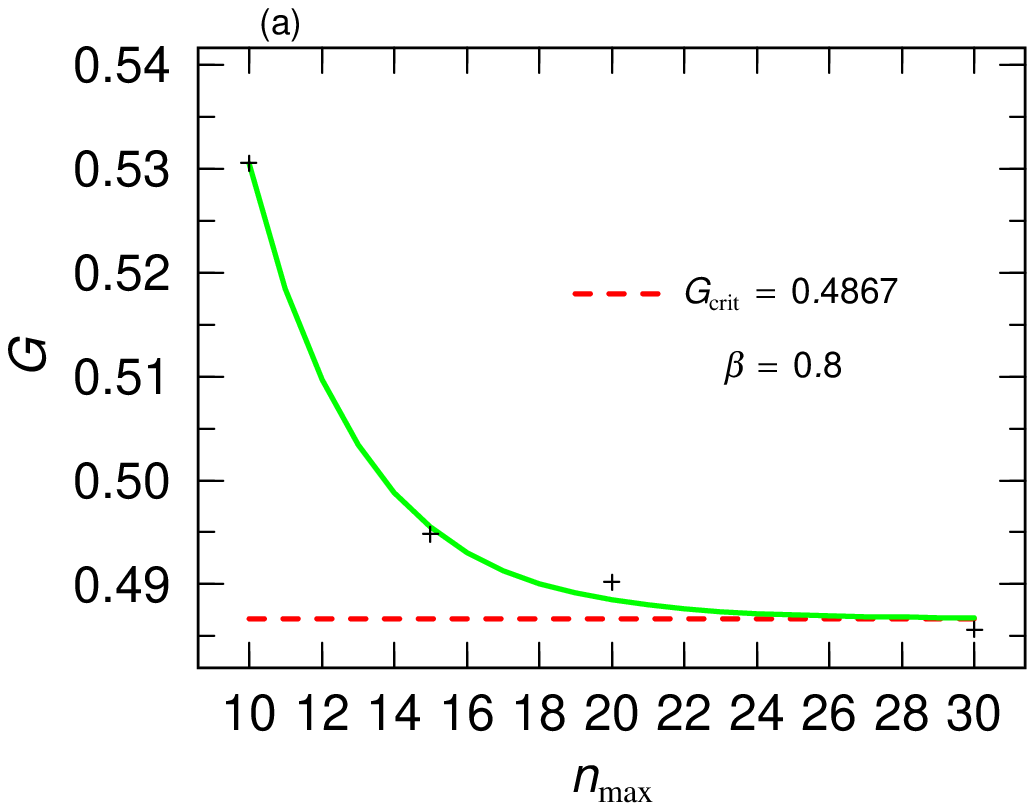}} \\
 \resizebox{85mm}{!}{\includegraphics{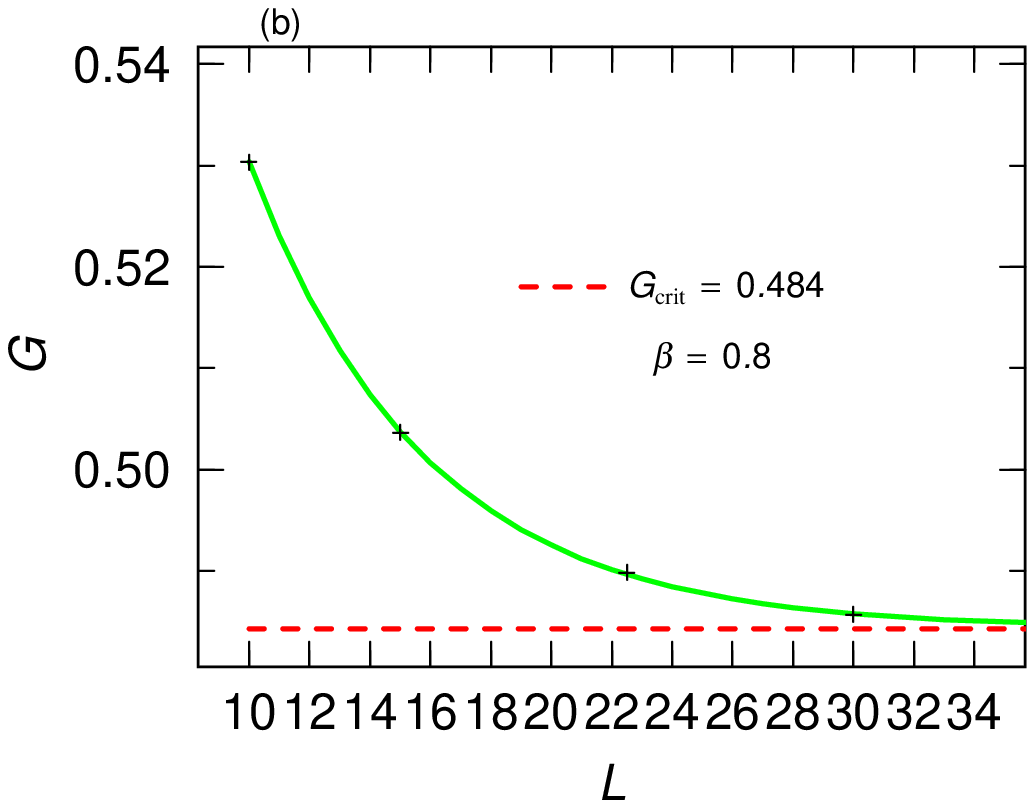}} 
    \end{tabular}
    \caption{(a) Numerical values of critical coupling strength $G$ 
as a function of $n_{\rm max}(=i_{\rm max})$ in the case of $\beta=0.8$ and $L(=2R)=30$. 
Cross symbols correspond to $n_{\rm max}=10,15,20,30$, respectively. 
Solid line is made by the function (\ref{modelA}), in which parameters $a,b,c$ are determined 
by the $\chi$-square fitting of those symbols; 
Dashed line corresponds to the parameter 
$a=G_{\rm crit}$ that is a critical coupling strength given by the extrapolation to $n_{\rm max}\rightarrow \infty$. 
(b) Numerical values of critical coupling strength $G$ 
as a function of $L(=2R)$ in the case of $\beta=0.8$ and $n_{\rm max}(=i_{\rm max})=30$. 
Symbols correspond to $L=10,15,22.5,30$, respectively. 
Solid line is the $\chi$-square fitting of these symbols using the function (\ref{modelA}) again.  
Dashed line corresponds to the parameter $a=G_{\rm crit}$:  
a deduced critical coupling strength for the spatially uniform system, i.e, at $L(=2R) \rightarrow \infty$.}
    \label{figLn}
  \end{center}
\end{figure}
\item
Then, in Fig.~\ref{figLn}(b) we show 
$G_{\rm crit}$ as a function of $L$ (symbols),  
and implement the extrapolation for $L(=2R) \rightarrow \infty$ 
using the same fitting function (\ref{modelA}) in order to 
deduce a critical coupling strength in spatially uniform system finally by $G_{\rm crit}=a$. 
\item
Subsequently, we repeat the same procedure implemented above  
but for different values of $\beta=1.0, 0.9, 0.7$ to obtain corresponding 
critical coupling strengths in spatially uniform system, 
which are summarized in Table~\ref{critG1}. 
\end{itemize}

\section{Validity condition of the Born approximation} 
\label{app:E}
In this appendix we evaluate the validity condition of the Born approximation 
for the RKKY potential in the spherical case, i.e., $\beta=1$. 
The condition demands that in the scattering processes the initial plane wave gives 
a primary contribution and the next leading order is negligible, which is rendered into 
\beq
1&\gg&  
\frac{2m_{22}}{4\pi} \left|
\int \rd^3\bx
\frac{e^{ikr}}{r} V(\bx) e^{i\bk \cdot \bx}
\right|
\nn
&=&
\frac{2m_{22}}{4\pi k_F^2}\left| 
V_0 \frac{4\pi}{\tilde{k}} \int_{0}^\infty \rd \tilde{r} \,  
e^{i\tilde{k}\tilde{r}}
\lk \frac{2\cos2\tilde{r}}{\tilde{r}^3} -\frac{\sin2\tilde{r}}{\tilde{r}^4}  \rk 
\sin(\tilde{k}\tilde{r})
\right|
\nn
&=&
\frac{2m_{22}V_0 }{3k_F^2\tilde{k}} 
\left| \pi \tilde{k}\lk \tilde{k}^2-3\rk + i 2\tilde{k}\lk \tilde{k}^2-3\rk
{\rm arctanh}(\tilde{k}) 
\right.
\nn 
&& \qquad \qquad \left. -i 2\tilde{k}^2-i 2\log\lk 1-\tilde{k}^2\rk 
\right|,  (0<\tilde{k}<1)
\nn 
\eeq
where $\tilde{k}=k/k_F$, $\tilde{r}=r k_F$, and $V_0 =g^2\frac{m_1 k_F^4}{16\pi^3\lambda^2}$. 
For low energies $\tilde{k} \ll 1$,  
the condition reduces to 
\beq
1&\gg& \frac{2m_{22}V_0 \pi}{k_F^2}   + \mathcal{O}\lk \tilde{k}^2\rk
\quad 
\rightarrow  \quad 
\frac{1}{\pi} \gg G,  
\eeq
where  $G$ is the dimensionless coupling constant (\ref{dimensionlessG1}). 
The condition can also be expressed in terms of the scattering length as 
\beq
&&\frac{1}{\pi} \gg \frac{|a_s|k_F}{2}. 
\eeq

\end{document}